\documentclass[11pt]{article}
\usepackage{amsfonts}
\usepackage{amssymb}
\usepackage{amsmath}
\usepackage{amsthm}
\usepackage{epsfig}
\usepackage{color}
\usepackage{hyperref}
\usepackage{cite}

\newlength{\bredde}
\def\slash#1{\settowidth{\bredde}{$#1$}\ifmmode\,\raisebox{.15ex}{/}
\hspace*{-\bredde} #1\else$\,\raisebox{.15ex}{/}\hspace*{-\bredde} #1$\fi}
\newcommand{\be}{\begin{equation}}                 
\newcommand{\ee}{\end{equation}}
\newcommand{\bea}{\begin{eqnarray}}
\newcommand{\eea}{\end{eqnarray}}
\newcommand{\nn}{\nonumber}
\newcommand{\eins}{\leavevmode\hbox{\small1\kern-3.8pt\normalsize1}}
\newcommand{\e}{\mbox{e}}
\newcommand{\la}{\lambda}

\newcommand{\sect}[1]{\setcounter{equation}{0}\section{#1}}

\def\Tr{{\mbox{Tr\,}}}
\def\re{{\mbox{Re}}}
\def\im{{\mbox{Im}}}
\def\Sdet{{\mbox{Sdet}}}
\def\Str{{\mbox{Str}}}
\def\diag{{\mbox{diag}}}

\begin{document}
\topmargin -1.4cm
\oddsidemargin -0.8cm
\evensidemargin -0.8cm
\title{\Large{{\bf
Spectral correlation functions of the sum of two independent complex Wishart 
matrices with unequal covariances}}}

\vspace{1.5cm}

\author{~\\{\sc Gernot Akemann}\footnote{akemann@physik.uni-bielefeld.de}, {\sc Tomasz Checinski}\footnote{tchecinski@uni-bielefeld.de} and {\sc Mario Kieburg}\footnote{mkieburg@physik.uni-bielefeld.de}
\\~\\
Department of Physics,
Bielefeld University,\\
Postfach 100131,
D-33501 Bielefeld, Germany}

\date{}
\maketitle
\vfill

\begin{abstract}
We compute the spectral statistics of the sum $H$ of two independent complex Wishart matrices, each of which 
is correlated with a different covariance matrix. Random matrix theory enjoys many applications including sums and products of random matrices. Typically ensembles with correlations among the matrix elements are much more difficult to solve. Using a combination of supersymmetry, superbosonisation and bi-orthogonal functions we are able to determine all spectral $k$-point density correlation functions of $H$ for arbitrary matrix size $N$. In the half-degenerate case, when one of the covariance matrices is proportional to the identity, the recent results by Kumar for the joint eigenvalue distribution of $H$ serve as our starting point. In this case the ensemble has a bi-orthogonal structure and we explicitly determine its kernel, providing its exact solution for finite $N$. The kernel follows from computing the expectation value of a single characteristic polynomial. In the general non-degenerate case the generating function for the $k$-point resolvent is determined from a supersymmetric evaluation of the expectation value of $k$ ratios of characteristic polynomials. Numerical simulations illustrate our findings for the spectral density at finite $N$ and we also give indications how to do the asymptotic large-$N$ analysis.
\end{abstract}

\vfill

\thispagestyle{empty}
\newpage

\renewcommand{\thefootnote}{\arabic{footnote}}
\setcounter{footnote}{0}

\sect{Introduction}\label{sec:intro}

One of the main goals and achievements of random matrix theory 
is the quantification of noise in the spectral statistics of some given data or operator. 
It allows to make analytical predictions for typical fluctuations and 
was called ``a new kind of statistical mechanics''  in~\cite{Guhr:1998}. One of the ensembles of random matrices most studied in physics and mathematics was introduced by Wishart~\cite{Wishart} in the context of mathematical statistics 
and is frequently used in multivariate statistics~\cite{Muirhead}.
Probably the best known quantity of the Wishart ensemble is its global spectral density derived by Marchenko and Pastur~\cite{MPlaw}, which is obtained in the limit of infinite matrix size. This and other more refined quantities like eigenvalue correlation functions or distributions of individual eigenvalues have found applications in physics, e.g. in Quantum Chromodynamics~\cite{Shuryak} and in other disciplines such as finance~\cite{finance}, medicine~\cite{medicine} climate research~\cite{climate}, telecommunication~\cite{telecommunication}, cf.~\cite{handbook} for a comprehensive list of modern applications.

In the classical Wishart ensemble a minimum of information is imposed by assuming that all its matrix elements are independent real or complex normal random variables. Due to this pro\-per\-ty this ensemble is exactly solvable for finite matrix size, see e.g.~\cite{Peter}. However, in realistic data system specific correlations among matrix elements are observed. In order to implement these the correlated Wishart ensemble has been introduced~\cite{Muirhead}. Here the eigenvalues and eigenvectors become coupled in a non-trivial way. Real matrices pose a considerable challenge as the group integrals involved in the calculation are not available. Nevertheless the spectral density~\cite{RKG10} has been computed exactly in this setting for finite matrices. For complex matrices the exact solvability persists, see~\cite{one-correl}.

Further generalisations that encode more structure of the observed correlations quickly lead to more difficult random one- or multi-matrix models. This is why we stick to complex matrices. The model introduced by Kumar~\cite{Kumar:2014} that we will study and solve in the present work belongs to this class. Let us now explain under which assumptions it arises. Consider an $N\times N_W$ matrix $W_{st}$ that contains $N$ different time series in its rows, measured at $N_W$ times steps given by its columns. When averaging over different realisations as denoted by brackets $\langle\ldots\rangle$ the most general correlations between two time steps $t$ and $t'$ and two time series $s$ and $s'$ read $\langle W_{st} W_{s't'}\rangle=\Sigma_{ss',tt'}$. Several approximations to this situation have been proposed. First, it was assumed that correlations in time steps and among time series factorise, as was considered in economics~\cite{economics}, climate research~\cite{climate}, sociology~\cite{sociology} and telecommunication~\cite{telecommunication}. In random matrix theory this problem was analytically discussed recently in~\cite{doubly-correl,WWG15} also for the real case and is called doubly-correlated Wishart ensemble. Second, a further strong simplification is to assume that different times steps are uncorrelated. Here analytical derivations for the spectral density were obtained in~\cite{RKG10,one-correl,RKGZ12}. In this simplified case also cross-correlations were considered~\cite{Vin1,Vin2}, in particular when the correlations among time series exhibit a matrix block-form, see also~\cite{time-lagged} for related time-lagged correlations.

In a very recent work by Kumar~\cite{Kumar:2014} the assumed absence of correlations among different time steps was softened. Two epochs of $N_A$ and $N_B$ time steps with $N_A+N_B=N_W$ were introduced, with different spatial correlations during the two epochs, see also~\cite{James} for an earlier work where such ensembles occur. This is equivalent to study the statistics of an $N\times N$ random matrix $H$ which is given by the sum of two correlated Wishart matrices, or for more epochs $T\geq2$ by $H=\sum_{j=1}^TA_jA_j^\dagger$. For two epochs, $T=2$, Kumar determined the joint density of eigenvalues of $H$ in terms of a hypergeometric function of matrix argument. In the case when one of the two correlation matrices is proportional to the identity matrix, called half-degenerate case, he showed that this joint density reduces to a bi-orthogonal ensemble containing ordinary confluent hypergeometric functions of Kummer type.  In this case he gave determinantal expressions of size $N+k$ containing moments of Kummer's hypergeometric functions for density with $k=1$~\cite{Kumar:2014} and for the $k$-point eigenvalue correlation functions~\cite{Kumar:2015}. Only in the completely degenerate case of two equal covariance matrices $H$ reduces to a single rectangular correlated Wishart matrix, where the spectral statistics are well known.

Our goal in the present work is to first exploit  the bi-orthogonal integrable structure of this ensemble at half-degeneracy in the sense of~\cite{Borodin:1998}, which was only mentioned in~\cite{Kumar:2014}. By explicitly determining the kernel of bi-orthogonal functions we provide the solution for all $k$-point correlation functions in terms of a determinant of size $k$ for arbitrary $N$. In this form the exact solution is amenable to study the asymptotic large-$N$ limit. Second, we aim at the solution for the general non-degenerate case that was not considered in~\cite{Kumar:2014,Kumar:2015}. Applying supersymmetric techniques~\cite{Efetov,Zirn06,Guhr} together with superbosonisation, we provide a closed form for the generating function of the $k$-point resolvent.

One particular application, where the solution of this ensemble is relevant for finite but not nec\-es\-sar\-i\-ly small $N$, is in the area of wireless telecommunication, in so-called multiple-input-multiple-output systems (MIMO)~\cite{Winters,Foschini,Telatar}. In the setup of $T=2$ the sum of two correlated Wishart matrices corresponds to the situation of two sets of transmitters and one set of receivers, with an obvious generalisation for larger $T$. The central quantity studied in MIMO systems is the mutual information averaged over different realisations called ergodic capacity. This quantity can be expressed as the expectation value of ${\ln}\left[{\det}\left[\eins +H\right]\right]$. Its generating function is closely related to those of the $k$-point resolvents that we compute.

Our work is organised as follows. In section~\ref{sec:problem} the ensemble we study is introduced and previous results from~\cite{Kumar:2014} and~\cite{Kumar:2015} are summarised. In section~\ref{sec:sol-hel-deg} we present our solution for the half-degenerate case. This includes a closed form expression for the partition function derived in appendix~\ref{sec:alter}, for the kernel in subsection \ref{sec:kern} and for the polynomials orthogonal to Kummer's hypergeometric function in subsection~\ref{sec:exp-single}. These polynomials are given by the expectation value of a single characteristic polynomial. We determine the latter for the general non-degenerate case. The expectation value of an inverse characteristic polynomial follows in subsection~\ref{sec:inv-single}. The spectral density in the half-degenerate case is given in detail in subsection~\ref{sec:density} and illustrated together with Monte Carlo simulations. The generating function for the $k$-point resolvent in the non-degenerate case is presented in section~\ref{sec:nondeg}, with details described in appendices~\ref{susyiso} and~\ref{sec:density-calc}. As an example we give an explicit representation for the spectral density and sketch its large-$N$ asymptotic analysis. At the end of section~\ref{sec:nondeg} further generalisations are briefly discussed. In section~\ref{sec:conclusio} we summarise our results and present an outlook.

\sect{Formulation of the Problem}\label{sec:problem}

In this section we define the random matrix ensemble that we consider and summarise the results derived by Kumar~\cite{Kumar:2014,Kumar:2015}, which serve as our starting point. Our notation follows closely the one employed in~\cite{Kumar:2014}.

We begin with two independent copies of complex correlated Wishart ensembles composed of Gaussian random matrices $A$ and $B$ of dimensions $N\times N_A$ and $N\times N_B$, respectively. Their dimensions are restricted to $N_A\geq N$ and $N_B\geq N$, and they are distributed as
\bea
\mathcal{P}_{A}\left(A\right)&=&\pi^{-NN_A}\det[\Sigma_{A}]^{-N_A}
\e^{-\Tr\left(\Sigma_{A}^{-1}AA^{\dagger}\right)}\,,\nn\\
\mathcal{P}_{B}\left(B\right)&=&\pi^{-NN_B}\det[\Sigma_{B}]^{-N_B}
\e^{-\Tr\left(\Sigma_{B}^{-1}BB^{\dagger}\right)}\,.\label{ProbDist}
\label{PABdef}
\eea
The probability distributions (\ref{ProbDist}) represent two independent complex correlated Wishart ensembles $AA^{\dagger}$ and $BB^{\dagger}$, respectively. Furthermore we require throughout our discussion that the fixed correlation matrices $\Sigma_A\neq \Sigma_B$ are positive definite, which should be by definition the case. The partition function is normalised to unity and it completely factorises,
\be
{\cal Z}_N=\int [dA] \int [dB]\  \mathcal{P}_{A}\left(A\right)\mathcal{P}_{B}\left(B\right)=1\,.
\ee
Integration measures over random matrices denoted by $[dX]$ are the Lebesgue measure meaning the product over the differentials of all independent real and imaginary parts of the elements of $X$.

We are interested in the spectral statistics of the sum of the two independent Wishart matrices,
\be
H\equiv AA^{\dagger}+BB^{\dagger}=WW^\dag,
\label{HWdef}
\ee
which is $N\times N$ Hermitian and positive definite. Here we have introduced the larger rectangular matrix $W=(A,B)$ of size $N\times(N_A+N_B)$. It consists of the elements $W_{i,j}=A_{i,j}$  and $W_{i,N_{A}+k}=B_{i,k}$ for $1\leq i\leq N$,  $1\leq j\leq N_A$ and $1\leq k\leq N_B$. This definition will be helpful later on due to the duality of the matrices $WW^{\dagger}$ and $W^{\dagger}W$, meaning that both matrices share the same non-zero eigenvalues.

In~\cite{Kumar:2014} two equivalent representations for the distribution of matrix elements of $H$ were shown to hold in the general case, valid for arbitrary non-degenerate correlation matrices $\Sigma_A$ and $\Sigma_B$: 
\bea
\mathcal{P}_{H}(H)&=&C_H\det[H]^{m}\e^{-\Tr(\Sigma_{A}^{-1}H)}\,_{1}\mathcal{F}_{1}\left(N_{B};N_{A}+N_{B};\left(\Sigma_{A}^{-1}-\Sigma_{B}^{-1}\right)H\right)
\nonumber\\ 
&=&C_H\det[H]^{m}\e^{-\Tr(\Sigma_{B}^{-1}H)}\,_{1}\mathcal{F}_{1}\left(N_{A};N_{A}+N_{B};\left(\Sigma_{B}^{-1}-\Sigma_{A}^{-1}\right)H\right)\label{Hyper1}\,.
\label{Hjpdf}
\eea
They are given in terms of the confluent hypergeometric function $_1\mathcal{F}_1$ of matrix argument, cf.~\cite{NIST} for a definition, together with
\bea
m&\equiv& N_{A}+N_{B}-N\,,
\label{mdef}\\
C^{-1}_H&\equiv&
\pi^{N(N-1)/2}\det[\Sigma_{A}]^{N_A}\det[\Sigma_{B}]^{N_B}\prod_{j=1}^N
\Gamma(m+j)\,.
\eea
For later use let us also define expectation values in the general non-degenerate case,
\be
\big\langle {\cal O}(H)\big\rangle_{N,N_A,N_B}^{\Sigma_A,\Sigma_B} \equiv \int [dH] 
 {\cal O}(H){\cal P}_H(H)\,.
\label{vevdef}
\ee
Here, ${\cal O}(H)$ is a function of the matrix $H$, e.g. the characteristic polynomial $\det\left[x\eins_N-H\right]$ that plays an important role later.

In the half-degenerate case where one of the covariance matrices is proportional to the identity, i.e. $\Sigma_{A}=\sigma_A\eins_N$ and $\Sigma_B=\mbox{diag}(\sigma_{B1},\ldots,\sigma_{B\,\!N})$, it was shown in~\cite{Kumar:2014} that the hypergeometric function in eq.~\eqref{Hjpdf} reduces to a determinant of Kummer's confluent hypergeometric functions $_{1}F_{1}$, using an identity from~\cite{Orlov:2004}. The joint probability density of eigenvalues $\la_j$, $j=1,\ldots,N$, of $H$ can then be written as 
\bea
P_N(\la_1,\ldots,\la_N)&=& C_{N,N_A,N_B}^{\Sigma_A,\Sigma_{B}}\prod_{j=1}^N\la_j^m\e^{-\la_j/\sigma_A}
\ \Delta_N(\{\la_i\})\nn\\
&&\times 
\det\left[\,_{1}F_{1}\left(m+1-N_A;m+1;
(\sigma_A^{-1}-\sigma_{Bk}^{-1})
\lambda_{l}\right)|_{1\leq k,l\leq N}\right] .
\label{jpdf}
\eea
Furthermore, in~\cite{Kumar:2014} an expression for $_1F_1$ at the above parameter values was given as a sum over incomplete Gamma-functions. In eq. (\ref{jpdf}) we also have introduced the Vandermonde determinant
\be
\Delta_N(\{\la_i\})\equiv \prod_{1\leq i<j\leq N} (\la_j-\la_i)=
\det[\la_i^{j-1}|_{1\leq i,j\leq N}]\,.
\label{Vandermonde}
\ee
Following~\cite{Kumar:2014}, the normalisation constant in eq.~(\ref{jpdf}) is given in terms of a determinant of Gauss' hypergeometric function $_{2}F_{1}$,
\bea
(C_{N,N_A,N_B}^{\Sigma_A,\Sigma_{B}})^{-1}&=&N!\,\sigma_A^{N m+\frac{N(N+1)}{2}}\prod_{k=1}^{N}\Gamma(m+k)\label{norm}\\
&&\times\det[\,_{2}F_{1}\left(m+1-N_A,m+1+j;m+1;
(\sigma_A^{-1}-\sigma_{Bi}^{-1})
\sigma_A\right)|_{1\leq i,j\leq N}]\,.\nn
\eea

In the completely degenerate limit $\sigma_{Bj}\to\sigma_A$ for all $j$, the differences $(\sigma_A^{-1}-\sigma_{Bi}^{-1})$ and thus both 
determinants in the normalised joint probability density, eq.~(\ref{jpdf}) with  eq.~(\ref{norm}), vanish. L'H\^opital's rule eventually reduces the limiting expression to the joint probability density of a single uncorrelated Wishart ensemble, which is also called Laguerre or chiral Gaussian unitary ensemble,
\be
\lim_{\sigma_{Bj}\to\sigma_A}P_N(\la_1,\ldots,\la_N)\sim \prod_{j=1}^N\la_j^m\e^{-\la_j/\sigma_A}
\ \Delta_N(\{\la_i\})^2\,.
\label{jpdfWL}
\ee
This classical ensemble can be solved in terms of generalised Laguerre polynomials as orthogonal polynomials, thus its name.

More generally it can be easily seen that in the limiting case of equal correlation matrices, $\Sigma_A=\Sigma_B=\Sigma$, not necessarily proportional to the identity 
matrix, the joint probability density of $H$ reduces to
\be
{\cal P}_H(H) \sim \det[H]^m\e^{-\Tr(\Sigma^{-1}H)}\,,
\label{degenerate}
\ee
corresponding to a single rectangular correlated Wishart ensemble.

We underline that the completely different limit, where $\sigma_{Bj}\to\sigma_B\neq\sigma_A$ for all $j$, does not yield the Laguerre ensemble as immediately follows from the representation~\eqref{jpdf-funct} of Kummer's confluent hypergeometric function. Then, the one-point weights resulting from eq. (\ref{PABdef}) consist of two different exponentials, namely $\exp[-\sigma_A^{-1}\lambda_j]$ and $\exp[-\sigma_B^{-1}\lambda_j]$, instead of one.

As it was mentioned in~\cite{Kumar:2014} the joint probability density~(\ref{jpdf}) in the half-degenerate case represents a bi-orthogonal ensemble in the sense of Borodin~\cite{Borodin:1998}. Consequently it satisfies a determinantal point process. Following Mehta~\cite{Mehta:2004} the $k$-point density correlation functions are defined by
\be
R_{k}\left(\la_{1},\ldots,\la_{k}\right)=
\frac{N!}{\left(N-k\right)!}
\prod_{j=k+1}^{N}\int_{0}^{\infty}d\lambda_{j} P_N\left(\lambda_{1},\ldots,\lambda_{N}\right)
\label{Rkdef1}
\ee
and can be expressed in terms of a single kernel~\cite{Borodin:1998}, once the bi-orthogonal functions are determined. For the proof of existence and equivalence of general systems of bi-orthogonal functions see \cite{Pell}. We will pursue this approach in section~\ref{sec:sol-hel-deg} below.

Following a different approach, in~\cite{Kumar:2014} in the half-degenerate case the spectral density $R_1(\lambda_1)$ of $H$ was expressed as a determinant of an $(N+1)\times(N+1)$ matrix by using the generalisation of the Andr\'eief integral formula, see~\cite{Kieburg:2010} or a summarised version in our appendix~\ref{Aint}. With the help of the same formula this was then generalised in~\cite{Kumar:2015} to the $k$-point density correlation function expressing it as a determinant of an $(N+k)\times(N+k)$ matrix:
\be
R_{k}\left(\la_{1},\ldots,\la_{k}\right)
=(-1)^{k}C_{N,N_A,N_B}^{\Sigma_A,\Sigma_{B}}N!\det\left[\begin{array}{rc}
\left.\int_0^\infty d\lambda \lambda^{j-1}\overset{\,}{\varphi}_{i}\left(\lambda\right)
\right|_{i=1,\ldots,N}^{j=1,\ldots,N}
&\left.\la_{i}^{j-1}\right|_{i=1,\ldots,k}^{j=1,\ldots,N}\\
\left.\overset{\,}{\varphi}_{j}\left(\la_{i}\right)
\right|_{i=1,\ldots,k}^{j=1,\ldots,N}
&{\mathbf 0}_{k\times k}\end{array}\right]\label{Andreief2}\,,
\ee
where
\begin{align}
\varphi_j(\la)&= \la^m \e^{-\la/\sigma_A}
\,_{1}F_{1}\left(m+1-N_A;m+1;
(\sigma_A^{-1}-\sigma_{Bj}^{-1})
\la \right)\ ,
\label{varphi}
\end{align}
with the normalisation constant given by (\ref{norm}). Whilst these representations are valid expressions, that may be useful for small matrix size $N$, they are clearly not suitable to take the large-$N$ limit or to study the issue of universality. In particular they do not exploit the integrable structure of a bi-orthogonal ensemble~\cite{Borodin:1998}, expressing the $k$-point correlation function through a determinant of size $k\times k$ of a single kernel. Our goal is the computation of this kernel.

\sect{Solution of the Half-Degenerate Case}\label{sec:sol-hel-deg}

In this section we present the solution of the simpler ensemble (\ref{jpdf}), i.e. in the half-degenerate case with $\Sigma_A=\sigma_A\eins_N$. In Section~\ref{sec:nondeg} we will discuss the more general non-degenerate case.

Let us first restate the starting point for our problem, the joint probability density (\ref{jpdf}):
\be
P_N(\la_1,\ldots,\la_N)\equiv C_{N,N_A,N_B}^{\Sigma_A,\Sigma_{B}}
\det[\la_i^{j-1}|_{1\leq i,j\leq N}]\ 
\det[\varphi_k(\la_l)|_{1\leq k,l\leq N}]\,.
\label{jpdf-biop}
\ee
Here we have simply used the second form of the Vandermonde determinant~(\ref{Vandermonde}) and moved the Laguerre weight factors $\la^m \e^{-\la/\sigma_A}$ in  eq.~(\ref{jpdf}) into the rows of the second determinant comprising $_1F_1$, yielding the one-point weight $\varphi_j$ defined in eq.~\eqref{varphi}. 

The building blocks in eq.~(\ref{jpdf-biop}) can be simplified further. In appendix~\ref{sec:alter}, where an alternative derivation for the joint density eq.~(\ref{jpdf}) is presented, an identity for the confluent hypergeometric function $_1F_1$ is derived. With the help of this identity the one-point weight $\varphi_j$ from eq.~\eqref{varphi} can be expressed for these parameter values in terms of elementary functions:
\begin{eqnarray}
\varphi_j(\la)  
&=&\exp[-\sigma_A^{-1}\lambda]\sum_{k=0}^{N_A-1}(-1)^k\frac{(N_A+N_B-N)!(N_B-N+k)!}{k!(N_A-1-k)!(N_B-N)!}\frac{\lambda^{N_A-1-k}}{(\sigma_{Bj}^{-1}-\sigma_A^{-1})^{N_B-N+1+k}}\nonumber\\ &&+\exp[-\sigma_{Bj}^{-1}\lambda]\sum_{k=0}^{N_B-N}(-1)^k\frac{(N_A+N_B-N)!(N_A-1+k)!}{k!(N_B-N-k)!(N_A-1)!}\frac{\lambda^{N_B-N-k}}{(\sigma_A^{-1}-\sigma_{Bj}^{-1})^{N_A+k}}\,.
\label{phi}
\end{eqnarray}
This relation also follows from the handbook on special functions~\cite{NIST} as explained in appendix~\ref{sec:alter}. It has to be compared with the expression as a sum over incomplete Gamma-functions given in \cite{Kumar:2014}, that simplifies to eq.~(\ref{phi}) only for $N_B=N$. More importantly, in appendix~\ref{sec:alter} the normalisation constant is found in a closed form in comparison to the determinant of $_2F_1$ in eq.~(\ref{norm}):
\begin{equation}\label{jpdf-const}
C_{N,N_A,N_B}^{\Sigma_A,\Sigma_{B}}=
\frac{\sigma_{A}^{-N_{A}N}\prod\limits_{k=1}^{N}\sigma_{Bk}^{N-N_B-1}}{N!\Delta_N(\{\sigma_{Bj}\})}\left(\prod_{l=0}^{N-1}\frac{(N_B-N)!}{(N_B-N+l)!(N_A+N_B-N)!}\right).
\end{equation}

Because of the bi-orthogonal structure of eq.~(\ref{jpdf-biop}) it is well known, see e.g.~\cite{Borodin:1998}, that the $k$-point correlation functions defined in eq.~(\ref{Rkdef1}) can be expressed in terms of a $k\times k$ determinant,
\be
R_{k}\left(\la_{1},\ldots,\la_{k}\right)=\det\left[ K_N(\la_i,\la_j)|_{1\leq i,j\leq k}\right]\,.
\label{Rkdet}
\ee
For $k=N$ we obtain an expression for the joint probability density which underlines the fact that this ensemble represents a determinantal point process. The kernel $K_N(\la,\mu)$ in eq. (\ref{Rkdet}) is in general given in terms of bi-orthogonal functions,
\be
\label{Ker1}
K_N(\la,\mu)=\sum_{j=0}^{N-1} \psi_j(\la)\phi_j(\mu)\ ,\ \ \mbox{with}
\ \ \delta_{jk}= \int_0^\infty d\la \psi_j(\la)\phi_k(\la)\ .
\ee
Here $\delta_{jk}$ is the Kronecker delta. The $\phi_j(\la)$ are in the linear span of the set $\{\varphi_l(\la),l=0,\ldots,j\}$ and the $\psi_j(\la)$ are polynomials of degree $j$ and thus in the linear span of the monomials inside the Vandermonde determinant~(\ref{Vandermonde}).

In our case the set $\{\varphi_l(\la)\}$ is special in the following sense. The functions defined in eq.~(\ref{varphi}) are the same functions in $\la$ for all $j$ and differ only through the argument $(\sigma_A^{-1}-\sigma_{Bj}^{-1})$. Thus in the following we construct polynomials orthogonal to the same function for different parameters $(\sigma_A^{-1}-\sigma_{Bj}^{-1})$. This will be done by expressing them through the expectation value of a single characteristic polynomial of $H=AA^\dagger+BB^\dagger$ in subsection~\ref{sec:exp-single}, with the help of the supersymmetry method~\cite{Efetov,Zirn06,Guhr}. The map to this expectation value is done by using the notion of the inverse Gram matrix in the next subsection~\ref{sec:kern}. After computing also the expectation value of an inverse characteristic polynomial in subsection~\ref{sec:inv-single} we conclude this section by discussing the result for the spectral density in subsection~\ref{sec:density}. There it is illustrated and compared with Monte Carlo simulations for two examples.

\subsection{The \texorpdfstring{$k$}{k}-point density correlation functions and their kernel}\label{sec:kern}

In this subsection we use an alternative form of the kernel $K_N(\la,\mu)$ compared to eq.~(\ref{Ker1}), which follows from~\cite{Borodin:1998}. There the following form for the kernel was derived for general bi-orthogonal ensembles:
\be
K_N(x,y)\equiv \sum_{i,j=1}^N x^{i-1} \left(g^{-1}\right)_{ij}\varphi_j(y)\,.
\label{kernel1}
\ee
It contains the inverse of the Gram matrix defined as
\be
g_{ij}\equiv \int_{0}^{\infty}d\lambda \lambda^{i-1}\varphi_{j}(\lambda)
=\sigma_A^{m+i}\Gamma(m+i)\,_{2}F_{1}\left(m+1-N_A,m+i;m+1;
(\sigma_A^{-1}-\sigma_{Bj}^{-1})
\sigma_A\right)\,,
\label{Gram}
\ee
where the second equality follows from an elementary integral. In~\cite{Kumar:2014} the normalisation constant in the form~(\ref{norm}) was expressed through the determinant of this Gram matrix, $(C_{N,N_A,N_B}^{\Sigma_A,\Sigma_{B}})^{-1}=N!\det[g]$, using the standard Andr\'eief formula~\cite{Andr}, see eq.~(\ref{genAnd}) with $k=l=0$.

Our task here is to map the problem of inverting the Gram matrix, see eq.~(\ref{kernel1}), to the evaluation of the expectation value of a single characteristic polynomial, yielding orthogonal polynomials in the sense of eq.~(\ref{Ker1}).

It is instructive to rederive the determinantal expression for the $k$-point correlation function~\eqref{Rkdet}, as it will help us to evaluate the kernel. Following the generalised Andr\'eief formula, see~\cite{Kieburg:2010} and our appendix~\ref{Aint}, the $k$-point density correlation functions can be written as an $(N+k)\times(N+k)$ determinant~(\ref{Andreief2}). Leaning on this determinantal expression we use the following identity for block determinants,
\be
\det\left[\begin{matrix}a&d\\c&b\end{matrix}\right]
=\det\left[a\right]{\rm det}\left[b-c\,a^{-1}d\right]\,,
\label{detid}
\ee 
where $a,b,c$ and $d$ are matrices with $a$ invertible. Identifying $b={\mathbf 0}_{k\times k}$ as the $k\times k$ matrix with zero in each matrix entry in the identity (\ref{detid}), we find the form of the kernel~(\ref{kernel1})
as a consequence.

Moreover, using the same line of computation backwards we obtain for the kernel:
\bea
K_N(x,y)&=&-C_{N,N_A,N_B}^{\Sigma_A,\Sigma_{B}}N!\det\left[\begin{array}{rc}
\left.\int_0^\infty d\lambda \lambda^{j-1}\overset{\,}{\varphi}_{i}\left(\lambda\right)
\right|_{i=1,\ldots,N}^{j=1,\ldots,N}
&\left.x^{j-1}\right|_{j=1,\ldots,N}\\
\left.\overset{\,}{\varphi}_{j}\left(y\right)
\right|_{j=1,\ldots,N}
&0\end{array}\right]\nn\\
&=&  NC_{N,N_A,N_B}^{\Sigma_A,\Sigma_{B}} \prod_{l=2}^{N}\int_{0}^{\infty}d\lambda_{l}\,
\det\left[
\begin{array}{c}
x^{j-1}\big|^{1\leq j\leq N}\\
\la_i^{j-1}\big|_{2\leq i\leq N}^{1\leq j\leq N}\\
\end{array}
\right]
\det\left[
\begin{array}{c}
\varphi_{j}(y)\big|^{1\leq j\leq N}\\
\varphi_{j}(\la_i)\big|_{2\leq i\leq N}^{1\leq j\leq N}\\
\end{array}
\right]\,.\label{kerneldet}
\eea
Expanding the first line with the help of the identity~(\ref{detid}), it becomes immediate that the right hand side is indeed the kernel~(\ref{kernel1}). From the second line in eq.~(\ref{kerneldet}), the well-known identification $K_N(x,x)=R_1(x)$ immediately follows.

Starting from the second line of eq.~(\ref{kerneldet}), we can use a simple Laplace expansion of the second determinant with respect to the first row containing the $\varphi_j(y)$. The $x$-dependence in the Vandermonde determinant can be rewritten as a smaller Vandermonde determinant in $\Lambda'={\rm diag}(\lambda_2,\ldots,\lambda_N)$, times a characteristic polynomial with $\Lambda'$ as a matrix and $x$ as its variable. Thus, we obtain
\bea
K_N(x,y)&=& N\,C_{N,N_A,N_B}^{\Sigma_A,\Sigma_{B}} \prod_{l=2}^{N}\int_{0}^{\infty}d\lambda_{l}\,
\sum_{j=1}^N(-1)^{j-1}\varphi_j(y) \det\left[\varphi_k(\la_l)\big|_{1\leq k\neq j\leq N}^{2\leq l\leq N}\right]\nn\\
&&\times\Delta_{N-1}(\{\lambda_2,\ldots,\lambda_N\})\prod_{k=2}^{N}(\la_k-x) \nn\\
&=& \sum_{j=1}^NG_j \big\langle\det[x\eins_{N-1}-H^\prime]\big\rangle_{N-1,N_A,N_B-1}^{\Sigma_A^\prime,\Sigma_{Bj}^\prime}\ \varphi_j(y)\,,
\label{kernel2}
\eea
where the functions $\varphi_{j}$ are still given by~(\ref{varphi}) or alternatively~(\ref{phi}) and the constants in the sum are defined by
\bea\label{pol-norm}
G_j\equiv (-1)^{N+j} N \frac{C_{N,N_A,N_B}^{\Sigma_A,\Sigma_{B}}}{C_{N-1,N_A,N_B-1}^{\Sigma_A^\prime,\Sigma_{Bj}^\prime}}=\frac{(N_B-N)!}{(N_B-1)!(N_A+N_B-N)!}\frac{1}{\sigma_A^{N_A}\sigma_{Bj}^{N_B-N+1}}\prod_{l\neq j}\frac{1}{(\sigma_{Bj}-\sigma_{Bl})}\,.
\eea
In the second equality of eq.~\eqref{kernel2} we have used that up to normalisation the integrals under the sum correspond to the expectation value of a single characteristic polynomial of an $(N-1)\times(N-1)$ random matrix $H^\prime$ with correlation matrices $\Sigma_A^\prime=\sigma_A\eins_{N-1}$ and $\Sigma_{Bj}^\prime=\mbox{diag}(\sigma_{B1},\ldots,\sigma_{Bj-1},\sigma_{Bj+1},\ldots,\sigma_{BN})$ where $\sigma_{Bj}$ is omitted. Note that also $N_B$ gets shifted to $N_B-1$ to guarantee that the parameters in the one-point weights $\varphi_j$ remain the same. In particular we use the fact that $N$ always appears in the combination $N_B-N$. The computation of the kernel is thus reduced to the computation of the  expectation value of a characteristic polynomial, with the expectation value defined in eq.~(\ref{vevdef}).

The kernel (\ref{kernel2}) together with eq.~(\ref{Rkdet}) is our first main result. In order to state the complete answer for the kernel we already give the result for the expectation value of the characteristic polynomial here:
\begin{align}
\hspace*{-0.4cm}\big\langle\det[x\eins_{N-1}-H^\prime]\big\rangle_{N-1,N_A,N_B-1}^{\Sigma_A^\prime,\Sigma_{Bj}^\prime}\ =N_{A}!(N_{B}-1)!\oint_{\gamma_{1}}\frac{dz_{1}}{2\pi i}
\oint_{\gamma_{2}}\frac{dz_{2}}{2\pi i}
\frac{\e^{z_{1}+z_{2}}}{z_{1}^{N_{A}+1}z_{2}^{N_{B}}}
\prod_{1\leq k\neq j \leq N}\left(x-z_{1}\,\sigma_{A}-z_{2}\,\sigma_{Bk}\right)\,.
\label{Doint}
\end{align}
The derivation of this quantity will be done in the next subsection~\ref{sec:exp-single} using supersymmetric methods

In the form~(\ref{kernel2}) the kernel is amenable to take the large-$N$ limit, replacing the sum by an integral and the summands given by the constants (\ref{pol-norm}), the double contour integral (\ref{Doint}) and the functions (\ref{varphi}) by their asymptotic values. The latter is known for general parameter values including large arguments and indices, see e.g. \cite{NIST}. 
Hence the large-$N$ limit of all correlation functions~(\ref{Rkdet}) can be derived in this way. 

Before coming to the expectation value of a single characteristic polynomial let us interpret the alternative representation of the kernel~(\ref{kernel2}) in comparison to eq.~(\ref{kernel1}). Typically the latter is simplified by making a change of basis of the following kind. Linear combinations within the linear span of the elements of the two determinants in eq. (\ref{jpdf-biop}) are sought, such that in the new basis the Gram matrix (\ref{Gram}) becomes diagonal and thus easy to invert. This construction reduces the double sum~(\ref{kernel1}) to a single sum over functions that are bi-orthogonal. It is also known that one of the two sets of functions, the bi-orthogonal polynomials, are spanned by the monic powers $\la_i^{j-1}$ and are given by the expectation value of a single characteristic polynomial. The difference between the standard literature and our case is that here all polynomials (\ref{Doint}) are of the same order, namely of order $N-1$, but nonetheless they build a basis. Usually one has for each order $0,1,2,\ldots$ one polynomial of this degree. The reason for this difference is due to the fact that all one-point weights $\varphi_j$ only differ by the parameter $\left(\sigma_A^{-1}-\sigma_{Bj}^{-1}\right)$ and nothing else. Keeping this basis $\{\varphi_j\}_{1\leq j\leq N}$ in the second determinant of the joint probability density~\eqref{jpdf-biop}, we have to get the same polynomial apart from an additional argument $\sigma_{Bj}$, due to symmetry reasons. Hence, our result~(\ref{kernel2}) is equivalent to build new polynomials of the same degree only from the linear span of the monomials  $\la_i^{j-1}$, while keeping the functions $\varphi_j(y)$ untouched. Consequently the two sets of functions satisfy the following orthogonality relation,
\be
\int_0^\infty dx 
P_{N-1}^{(j)}(x)
\varphi_k(x) = G_j^{-1}\delta_{jk}\,.
\label{biop}
\ee
Here we have introduced the polynomial $P_{N-1}^{(j)}(x)$ of degree $N-1$ in monic normalisation that depends on all $\sigma_{B i}$ with $i\neq j$,
\be
P_{N-1}^{(j)}(x)\equiv
\big\langle\det\left[x\eins_{N-1}-H^\prime\right]\big\rangle_{N-1,N_A,N_B-1}^{\Sigma_A^\prime,\Sigma_{Bj}^\prime}
\ =\ G_j^{-1}\sum_{i=1}^{N}x^{i-1}(g^{-1})_{ij}\,.
\label{OPdef}
\ee
This expression for the polynomials bi-orthogonal to the functions $\varphi_j$ resembles the Heine formula~\cite{Mehta:2004} for orthogonal polynomials. The normalisation constants $G_j$ appear in this relation due to the diagonalisation of the Gram matrix $g$.

Equation~(\ref{biop}) can be easily cross checked by explicitly writing the expectation value in terms of the defining integral. The integration over all variables $x,\la_2,\ldots,\la_N$ times the Vandermonde determinant anti-symmetrises in all variables. Whenever $j\neq k$ we will encounter the product $\varphi_k(\la_l)\varphi_k(x)$ in the Laplace expansion of the second determinant, which is symmetric in the two arguments $\la_l$ and $x$ and thus vanishes. 

\subsection{Expectation value of a single characteristic polynomial}\label{sec:exp-single}

We now compute the expectation value of the characteristic polynomial of degree $N$,
\be
P_N(x)\equiv \big\langle\det\left[x\eins_N-WW^\dag\right]\big\rangle_{N,N_A,N_B}^{\Sigma_A,\Sigma_{B}}\,.
\label{OP}
\ee
From this expression the polynomial $P_{N-1}^{(j)}(x)$ needed for the kernel~(\ref{kernel2}) simply follows by reducing $N\to N-1$, $N_{B}\to N_{B}-1$ and omitting the eigenvalue $\sigma_{Bj}$ of $\Sigma_B$. We directly consider the non-degenerate case, $\Sigma_A\neq\sigma_A\eins_N$, as it is not more complicated than in the
half-degenerated case. Moreover, the computation in this subsection sketches already the main ideas to be applied in section~\ref{sec:nondeg}, where the generating function for the $k$-point density correlation function 
is computed in the general case. This generating function is given by expectation values of $k$ ratios of characteristic polynomials.

Our derivation uses the duality between $WW^\dagger$ and $W^\dagger W$ in the first step, then we express the determinant through Gaussian integrals of fermionic nature and perform the Gaussian average over $W$. In a final step we make use of the superbosonisation formula~\cite{BEKYZ,Som07,BA07,LSZ08,KSG09} to reduce the remaining number of integrals to a minimum.

The duality between $WW^\dagger$ and $W^\dagger W$ implies that we can write eq.~(\ref{OP}) equivalently as
\be
P_N(x)=x^{N-N_W}\big\langle\det\left[x\eins_{N_W}-W^\dag W\right]\big\rangle_{N,N_A,N_B}^{\Sigma_A,\Sigma_{B}}\,,
\label{OP2}
\ee
as the matrix $WW^\dag$ of dimension $N$ and the matrix $W^\dag W$ of dimension $N_W=N_A+N_B$ have the same $N$ non-vanishing eigenvalues. In the case that $WW^\dag$ has full rank $N$ then $W^\dag W$ has $N_W-N$ zero eigenvalues.

It is well known that the determinant of a matrix can be expressed through an integral over Grassmann variables, also called Berezin integral. Let us introduce a set of $N_W$ complex (anti-commuting) Grassmann variables $v_i$, with the following convention for complex conjugation,
\be
\left\{v_{i},v_{j}\right\}=0\,,\ \{v_{i},v_{j}^{*}\}=0\,,\ \left(v_{i}^{*}\right)^{*}=-v_{i}\,,\quad\forall\, i,j=1,\ldots,N_W\,,
\label{Galgebra}
\ee
with $\left\{,\right\}$ representing the anti-commutator.

We arrange the $v_j$ in two sets of vectors,
\be
V\equiv\left(\begin{matrix}V_{A}\\V_{B}\end{matrix}\right),\quad V_{A}\equiv\left(\begin{matrix}v_{1}\\\vdots\\v_{N_{A}}\end{matrix}\right)\quad\text{and }\quad V_{B}\equiv\left(\begin{matrix}v_{N_{A}+1}\\\vdots\\v_{N_{W}}\end{matrix}\right),
\label{VABdef}
\ee
with scalar product
\be
V^{\dagger}V=\sum\limits_{i=1}^{N_{W}}v_{i}^{*}v_{i}\,.
\label{scalar}
\ee
The standard flat integration measure for Grassmann variables can be defined for one particular Grassmann variable $v_{i}$ by only two properties, with an obvious extension to multiple integrations:
\be
\int dv_{i}v_{i}= 1\,,\quad\int dv_{i}=0\,,\quad\forall\, i=1,\ldots , N_{W}\quad\text{and }\quad\left[dV\right]\equiv\prod_{i=1}^{N_{W}}dv_{i}^{*} dv_{i}\,.
\label{Gint}
\ee
Here the last definition is required because of a possible minus sign. With these definitions the characteristic polynomial~(\ref{OP2}) can be written as 
\bea
P_{N}(x)&=&x^{N-N_{W}}\int \left[dV\right]
\left\langle
\exp\left[-V^\dag(x\eins_{N_W} -W^\dag W)V\right]
\right\rangle_{N,N_A,N_B}^{\Sigma_A,\Sigma_{B}}
\nn\\
&=&x^{N-N_{W}}\det[\Sigma_A]^{-N_A}\det[\Sigma_B]^{-N_B}
\int \left[dV\right]\e^{-xV^{\dagger}V}\nn\\
&&\times
\det\left[
\begin{array}{cc}
\Sigma_{A}^{-1}\otimes\eins_{N_A}+\eins_N\otimes V_A^{}V_A^\dag
&\eins_N\otimes V_A^{}V_B^\dag\\
\eins_N\otimes V_B^{}V_A^\dag&
\Sigma_{B}^{-1}\otimes\eins_{N_B}+\eins_N\otimes V_B^{}V_B^\dag\\
\end{array}
\right]^{-1}.
\label{OPave}
\eea
In the second equality we have performed the average in $W$ according to the correlated Gaussian averages~(\ref{PABdef}). Note that the matrices $V_AV_A^\dag$, $V_BV_A^\dag$ etc. are matrices composed as a dyadic tensor by taking the dyadic product of the two vectors, which are bosonic objects. Hence its matrix entries are commuting.

In order to simplify the determinant in the integrand of the block-matrix
\begin{equation}
\label{DD}
\mathcal{D}=\left(\begin{matrix}a&d\\c&b\end{matrix}\right)=\left(
\begin{array}{cc}
\Sigma_{A}^{-1}\otimes\eins_{N_A}+\eins_N\otimes V_A^{}V_A^\dag
&\eins_N\otimes V_A^{}V_B^\dag\\
\eins_N\otimes V_B^{}V_A^\dag&
\Sigma_{B}^{-1}\otimes\eins_{N_B}+\eins_N\otimes V_B^{}V_B^\dag\\
\end{array}
\right),
\end{equation} 
we apply the identity $\det\mathcal{D}=\det[a]\det[b]\det[1-b^{-1}ca^{-1}d]$. The first two determinants are
\bea
\det[a]&=&\det\left[\Sigma_{A}^{-1}\otimes\eins_{N_{A}}
+\eins_{N}\otimes V_A^{}V_A^\dag\right]
=\det[\Sigma_{A}]^{-N_{A}}
\det\left[\eins_{N}+V_A^\dag V_A^{}\Sigma_{A}\right]^{-1} ,\nn\\
\det[b]&=&\det\left[\Sigma_{B}^{-1}\otimes\eins_{N_{B}}
+\eins_{N}\otimes V_B^{}V_B^\dag\right]
=\det[\Sigma_{B}]^{-N_{B}}
\det\left[\eins_{N}+V_B^\dag V_B^{}\Sigma_{B}\right]^{-1}.
\label{blockab}
\eea
In the second step we pulled the matrices $\Sigma_A$ and $\Sigma_B$ out of the determinants and used a well-known duality for Grassmann variables $\det[\eins_{N_A}+\gamma V_AV_A^\dag]=(1+\gamma V_A^\dag V_A)^{-1}$ with $\gamma$ an arbitrary constant. Note that $V_A^\dag V_A$ and $V_B^\dag V_B$ are scalars as mentioned in eq.~(\ref{scalar}). This fact allows to reduce the size of the determinants to $N$ instead of $NN_A$ and $NN_B$, respectively. Furthermore we may write in eq. (\ref{DD})
\bea
d&=&\eins_{N}\otimes V_{A}^{}V_{B}^{\dagger}=\big(\eins_{N}\otimes V_{A}^{}\big)\big(\eins_{N}\otimes V_{B}^{\dagger}\big)\,,\nn\\
c&=&\eins_{N}\otimes V_{B}^{}V_{A}^{\dagger}=\big(\eins_{N}\otimes V_{B}^{}\big)\big(\eins_{N}\otimes V_{A}^{\dagger}\big)\,
\label{Splittingcd} .
\eea
This becomes useful for the remaining determinant for which one can make the following expansion
\begin{equation}
\det[\eins-b^{-1}ca^{-1}d]\equiv\det[\eins-C]=\exp\left[-\sum_{k=1}^\infty\frac{\Tr(C)^k}{k}\right].
\end{equation}
The last factor in $C$ is $(\eins_{N}\otimes V_{B}^{\dagger})$ from the matrix $d$ which will be moved to the first position of $C$, using the cyclicity of the trace at the expense of a minus sign. The minus sign is due to the fermionic (anti-commuting) nature of $d$ and of $b^{-1}ca^{-1}$. This minus carries over to the inverse of the determinant $\det[\eins-\tilde{C}]^{-1}$, where $\tilde{C}=db^{-1}ca^{-1}$. We thus obtain
\bea
\det[\eins-b^{-1}ca^{-1}d]&=&\det\left[\eins_{N}\,
-(\eins_{N}\otimes V_B^\dag)
\left(\eins_{N}\otimes\eins_{N_{B}}+\Sigma_{B}\otimes V_B^{}V_B^\dag\right)^{-1}
(\Sigma_{B}\otimes V_{B}^{})\right.
\\\label{det3}
&&\quad\quad\quad\quad \times
\left.(\eins_{N}\otimes V_{A}^{\dagger})
\left(\eins_{N}\otimes\eins_{N_{A}}+\Sigma_{A}\otimes V_{A}^{}V_{A}^\dag\right)^{-1}
(\Sigma_{A}\otimes V_{A}^{})\right]^{-1}\,\nn\\
&=&\det\left[\eins_{N}\,
-\left(\eins_{N}+V_{B}^{\dagger}V_{B}^{}\Sigma_{B}\right)^{-1}
V_{B}^{\dagger}V_{B}^{}\Sigma_{B}
\left(\eins_{N}+V_{A}^{\dagger}V_{A}^{}\Sigma_{A}\right)^{-1}
V_{A}^{\dagger}V_{A}\Sigma_{A}\right]^{-1}.\nn
\eea
In the last step we have commuted the following two factors. Expanding the geometric series of the second factor into its von Neumann series we obtain
\bea
(\eins_{N}\otimes V_{B}^{\dagger})
\left(\eins_{N}\otimes\eins_{N_{B}}+\Sigma_{B}\otimes V_{B}^{}V_{B}^{\dagger}\right)^{-1}
&=&(\eins_{N}\otimes V_{B}^{\dagger})
\sum\limits_{k=0}^{\infty}(-1)^k(\Sigma_{B}\otimes V_{B}^{}V_{B}^{\dagger})^{k}
\nn\\
&=&
\sum\limits_{k=0}^{\infty}(-V_{B}^{\dagger}V_{B}^{}\Sigma_{B})^{k}
(\eins_{N}\otimes V_{B}^{\dagger})
\nn\\
&=& 
\left(\eins_{N}+V_{B}^{\dagger}V_{B}^{}\Sigma_{B}\right)^{-1}
(\eins_{N}\otimes V_{B}^{\dagger})\ ,
\eea
and likewise for the two factors  $(\eins_N\otimes V_{A})$ and $\left(\eins_{N}\otimes\eins_{N_{A}}+\Sigma_{A}\otimes V_{A}^{}V_{A}^\dag\right)^{-1}$ as in eq.~(\ref{det3}). We can now insert eqs.~(\ref{blockab}) and (\ref{det3}) into eq.~(\ref{OPave}) and obtain
\bea
P_{N}(x)&=&x^{N-N_{W}}\det[\Sigma_A]^{-N_A}\det[\Sigma_B]^{-N_B}\int \left[dV\right]\ \e^{-xV^{\dagger}V}
\det[a]^{-1}\det[b]^{-1}\det[\eins-b^{-1}ca^{-1}d]^{-1}
\nn\\
&=&x^{N-N_{W}}\int \left[dV\right] \ \e^{-xV^{\dagger}V}
\det\left[\eins_N+V_A^\dag V_A^{}\Sigma_{A}+V_B^\dag V_B^{}\Sigma_{B}\right] , 
\label{OPgen}
\eea
after cancelling all normalisation factors and combining the three determinants. This is the result for the expectation value of a characteristic polynomial valid for arbitrary covariance matrices $\Sigma_A$ and $\Sigma_B$.

Moreover, in the case when these two matrices commute and thus can be simultaneously diagonalised, which is in particular true for the half-degenerate case $\Sigma_A=\sigma_A\eins_N$, the result (\ref{OPgen}) can simply be expressed in terms of the two sets of eigenvalues $\{\sigma_{Ak}\}$ and $\{\sigma_{Bk}\}$ of the two matrices $\Sigma_A$ and $\Sigma_B$,
\be
P_{N}(x)=x^{N-N_{W}}\int \left[dV\right] \ \e^{-xV^{\dagger}V}
\prod_{k=1}^{N}\left(1+V_A^\dag V_A^{}\sigma_{Ak}+V_B^\dag V_B^{}\sigma_{Bk}\right)\,.
\label{OPshort}
\ee
In this case the determinant reduces to a simple product.

In the final step we apply the superbosonisation formula~\cite{BEKYZ,Som07,BA07,LSZ08,KSG09}. In the present case one can readily understand this formula via a Taylor expansion. For any entire function, $f(z)=\sum_{k=0}^\infty f^{(k)}(0)z^k/k!$, that only depends on the scalar product~(\ref{scalar}), $f=f(V^\dag V)$, its Grassmann integral can be efficiently evaluated. From eq.~(\ref{Gint}) it is clear that only the term containing all pairs $v_i^*v_i$ contributes. This term is contained only in the power $(V^\dag V)^{N_W}$, with multiplicity $N_W !$. Thus the Grassmann integral becomes in this case 
\be
\int d\left[V\right]f(V^{\dagger}V)=\int d\left[V\right]f^{(N_W)}(0)
\prod_{k=1}^{N_W}v_k^*v_k=(-1)^{N_W}N_W!\oint_{\gamma} \frac{dz}{2\pi i}\frac{1}{z^{N_W+1}}f(z)\ .
\label{sbos1}
\ee
The integration contour $\gamma$ encloses the origin in positive direction. This expression is the superbosonisation formula in the one-dimensional fermionic case. The only complication from eq.~(\ref{OPshort}) is that the scalar products $V_A^\dag V_A^{}$ and $V_B^\dag V_B^{}$ appear with different factors. Hence we apply eq.~(\ref{sbos1}) twice, for each of the two sets of variables $V_A$ and $V_B$. We arrive at
\be
P_{N}(x)=N_{A}!N_{B}!\oint_{\gamma_{1}}\frac{dz_{1}}{2\pi i}
\oint_{\gamma_{2}}\frac{dz_{2}}{2\pi i}
\frac{\e^{z_{1}+z_{2}}}{z_{1}^{N_{A}+1}z_{2}^{N_{B}+1}}
\det\left[x\eins_{N}-z_{1}\Sigma_A-z_{2}\Sigma_B\right]\,.
\label{OPfinal}
\ee
Note that we rescaled the contour integrals by the variable $-x$ to absorb the additional prefactor $x^{N-N_{W}}$ and the sign in the integrand.

Equation~\eqref{OPfinal} is the second main result of this section. The expectation value of a single characteristic polynomial reduces to this simple expression, valid for commuting correlation matrices $\Sigma_A$ and $\Sigma_B$ with $[\Sigma_A,\Sigma_B]=0$. From this the polynomial $P_{N-1}^{(j)}(x)$ in eq.~(\ref{OPdef}), trivially follows, by setting $\Sigma_A^\prime=\sigma_A\eins_{N-1}$\,,\;$\Sigma_B^\prime={\rm diag}(\sigma_{B1},\ldots,\sigma_{B,j-1},\sigma_{B,j+1},\ldots,\sigma_{BN})$ and by replacing $N\to N-1$ and $N_B\to N_B-1$ in the average. Then this polynomial reads
\be
P_{N-1}^{(j)}(x)=N_{A}!(N_{B}-1)!\oint_{\gamma_{1}}\frac{dz_{1}}{2\pi i}
\oint_{\gamma_{2}}\frac{dz_{2}}{2\pi i}
\frac{\e^{z_{1}+z_{2}}}{z_{1}^{N_{A}+1}z_{2}^{N_{B}}}
\prod_{1\leq k\neq j \leq N}\left(x-z_{1}\,\sigma_{A}-z_{2}\,\sigma_{Bk}\right)
\label{OPjfinal}
\ee
and together with eq.~(\ref{kernel2}) determines the kernel and thus all $k$-point correlation functions of this model.

We conclude this subsection with a consistency check. As we have pointed out earlier in eq.~(\ref{degenerate}) for equal correlation matrices our setting reduces to a single correlated Wishart-Laguerre ensemble of matrix dimensions $N\times N_W$. In the completely degenerate case $\Sigma_A=\Sigma_B=\sigma_{A}\eins_N$ it is well known that the expectation value of a characteristic polynomial is simply given by the generalised Laguerre polynomial orthogonal with respect to the weight function $w(x)=x^{N_{W}-N}\exp[-x/\sigma]$ in monic normalisation, cf. eq.~(\ref{jpdfWL}) for the joint density of all eigenvalues. Our result agrees with this limiting case as follows.

In eq.~(\ref{OPshort}) a single application of the superbosonisation formula~(\ref{sbos1}) suffices when setting all eigenvalues of the correlation matrices $\Sigma_{A}$ and $\Sigma_{B}$ to be equal, $\sigma_{Ak}=\sigma_{Bk}=\sigma$. From there we obtain for the fully degenerate case~(\ref{jpdfWL})
\be
P_{N}^{deg}(x)=N_W!\oint_{\gamma}\frac{dz}{2\pi i}
\frac{\e^{z}}{z^{N_W+1}} (x-z\,\sigma)^N\, .
\label{OPdeg}
\ee
This has to be compared to the standard complex contour integral representation of the generalised Laguerre polynomial, see e.g.~\cite{NIST}, 
\be
L_n^{\alpha}(x)=\oint_{{\cal C}}\frac{dz}{2\pi i}\frac{z^n}{(z-x)^{n+1}}
\frac{z^\alpha}{x^\alpha}\e^{-z+x}=\frac{(-1)^{n}}{\sigma^{n+\alpha}x^\alpha}\oint_{\gamma} \frac{dv}{2\pi i}\frac{(x\sigma-v\sigma)^{n+\alpha}}{v^{n+1}}\ \e^{v}\,.
\label{Lrep}
\ee
In the second equation we have simply shifted and rescaled the contour ${\cal C}$, formerly enclosing the point $x$ and not the origin, to a contour $\gamma$ around the origin (and not including $v=x$). After an appropriate rescaling we thus have 
\bea
P_{N}^{deg}(x)&=&(-\sigma)^{N_W}N_W! x^{N-N_W}L_{N_W}^{N-N_W}(x/\sigma)\nn\\
&=&(-\sigma)^{N_W}N_W! x^{N-N_W}\sum_{k=0}^{N_W}
\frac{N!}{(N_W-k)!\Gamma(k+N-N_W+1)k!}\left(-\frac{x}{\sigma}\right)^k\nn\\
&=&(-\sigma)^NN!\sum_{l=0}^{N}\frac{N_W!}{(N-l)!l!(N_W-N+l)!}\left(-\frac{x}{\sigma}\right)^l\nn\\
&=&(-\sigma)^NN!L_N^{N_{W}-N}(x/\sigma)\,.
\eea
In the first step we have used the explicit representation of the generalised Laguerre polynomial. Due to the fact that $N-N_W=N-N_A-N_B<0$, the Gamma-function in the denominator truncates the sum from below up to $k=N_W-N$. A shift $l=k-N_W+N$ leads to the desired form, the generalised Laguerre polynomial of degree $N$ in monic normalisation. It is orthogonal with respect to the weight function $x^{N_W-N}\e^{-\sigma^{-1}x}$, as desired for the limiting ensemble (\ref{jpdfWL}) with $\sigma=\sigma_{A}$.

\subsection{Expectation value of an inverse characteristic polynomial}\label{sec:inv-single}

Let us compute the expectation value of an inverse characteristic polynomial,
\be
Q_{N}(y)\equiv 
\left\langle\det[y\eins_{N}-WW^{\dagger}]^{-1}\right\rangle_{N,N_A,N_B}^{\Sigma_A,\Sigma_B}
=y^{-N}\left\langle\det[\eins_{N_{W}}-W^{\dagger}W/y]^{-1}\right\rangle_{N,N_A,N_B}^{\Sigma_A,\Sigma_B}.
\label{Qdef}
\ee
Here we have to choose $\im (y)\neq 0$ in order to regularise the expression. This quantity is related to the Cauchy-transform of the one-point weights $\varphi_j$. This can be seen by combining the Vandermonde determinant $\Delta_{N}(\{\lambda_j\})$ of the joint probability density~\eqref{jpdf-biop} with the inverse determinant in eq.~\eqref{Qdef}, i.e.
\begin{equation}
\frac{\Delta_{N}(\{\lambda_j\})}{\prod_{j=1}^N(y-\lambda_j)}=\det\left[\begin{array}{c} \lambda_j^{k-1}|^{1\leq k\leq N-1}_{1\leq j\leq N} \\ \displaystyle\left.\frac{1}{y-\lambda_j}\right|_{1\leq j\leq N}\end{array}\right],
\end{equation}
see~\cite{Kieburg:2010}. This determinant can be expanded in the last row yielding $N$ terms in eq.~\eqref{jpdf-biop}. All $N$ terms give the same integral such that
\begin{equation}
Q_{N}(y)=NC_{N,N_A,N_B}^{\Sigma_A,\Sigma_B}\prod_{j=1}^{N}\int_{0}^{\infty}d\lambda_{j}\frac{\Delta_{N-1}(\lambda_1,\ldots,\lambda_{N-1})}{y-\lambda_N}\det\left[\begin{array}{l} \varphi_{l}(\lambda_k)|_{1\leq l\leq N}^{1\leq k\leq N-1} \\ \varphi_{l}(\lambda_N)|_{1\leq l\leq N} \end{array}\right],
\end{equation}
Likewise we expand the second determinant in $\varphi_{l}\left(\lambda_N\right)$. The remaining integrals yield the constants $1/C_{N-1,N_A,N_B-1}^{\Sigma'_A,\Sigma'_{B,j}}$. Then we have
\begin{eqnarray}
Q_{N}(y)=\sum_{j=1}^N(-1)^{N+j}N\frac{C_{N,N_A,N_B}^{\Sigma_A,\Sigma_B}}{C_{N-1,N_A,N_B-1}^{\Sigma'_A,\Sigma'_{B,j}}}\int_{-\infty}^\infty\frac{dx}{y-x}\varphi_{j}(x)=\int_{-\infty}^\infty\frac{dx}{y-x}\sum_{j=1}^N G_j\varphi_{j}(x)\,.
\end{eqnarray}
Here, we used the definition~\eqref{pol-norm} for the constants $G_j$. Hence $Q_{N}(y)$ is the Cauchy transform of a weighted average of the functions $\varphi_j(x)$. Note that in the standard setting of bi-orthogonal polynomials $Q_{N}(y)$ would be  given instead by the Cauchy transform of a single orthogonal polynomial.

Let us come to the calculation of the average~\eqref{Qdef}. In this case the inverse determinant can be expressed as a Gaussian integral over two complex vectors $U_A$ and $U_B$ of ordinary bosonic variables, in analogy to eq.~(\ref{VABdef}). Apart from signs due to the fermionic nature of $V$ and the bosonic one of $U$ the computation is very similar to the one presented in the previous subsection. In contrast to the fermionic case we will encounter poles, which is why we have to keep track of the regulating imaginary part $\im (y)\neq0$.

After integrating over the Gaussian matrices $A$ and $B$, using the duality (\ref{Qdef}) and identities for determinants we arrive at 
\be
Q_N(y)=\frac{1}{\pi^{N_W}}\int \left[dU\right]\e^{-U^{\dagger}U}
\det\left[y\eins_N-U_{A}^{\dagger}U_{A}\Sigma_{A}-U_{B}^{\dagger}U_{B}\Sigma_{B}\right]^{-1}\label{QbeforeSB},
\ee
with $[dU]$ being the flat measure on ${\mathbb C}^{N_W}$. This is valid for arbitrary covariance matrices $\Sigma_A$ and $\Sigma_B$. In the case of commuting covariance matrices, $\left[\Sigma_{A},\Sigma_{B}\right]=0$, and in particular in the half-degenerate case the determinant in  eq.~(\ref{QbeforeSB}) can be diagonalised, and the integral simplifies further. The bosonisation formula for the present case reads
\be
\int d\left[U\right]f\left(U^{\dagger}U\right)=
\frac{\pi^{N_W}}{\left(N_W-1\right)!}\int\limits_{0}^{\infty}ds\ s^{N_W-1}f(s)\ .
\ee
This follows from going over to polar coordinates in $2N_W$ real dimensions for $U$, with the scalar product $U^\dag U=s$ being the squared norm of the complex vector $U$. Applied to both $U_A$ and to $U_B$ we find the final expression valid for commuting covariance matrices, compared to eq. (\ref{OPfinal}):
\be
Q_{N}(y)=\frac{\pi^{N_W}}{(N_{A}-1)!(N_{B}-1)!}
\int\limits_{0}^{\infty}ds_{1}\int\limits_{0}^{\infty}ds_{2}\ s_{1}^{N_{A}-1}s_{2}^{N_{B}-1}
\e^{-\left(s_{1}+s_{2}\right)}\det\left[y\eins_{N}-s_{1}\Sigma_{A}-s_{2}\Sigma_{B}\right]^{-1}.
\label{Qfinal}
\ee
The correct monic normalisation can be easily checked, by observing that $\lim_{|y|\to\infty}y^NQ_N(y)=1$, as it is required from the definition~(\ref{Qdef}).

\subsection{The spectral density}\label{sec:density}

In this subsection we will apply the previous results to the spectral density and give an alternative integral representation for the density and kernel. They are illustrated by numerical simulations below. 

For this purpose we combine the result for the orthogonal polynomial~\eqref{OPjfinal} with the expression for the kernel~\eqref{kernel2} including $\varphi_{j}$, see eq.~(\ref{varphi}). In particular we exchange the finite sum with the integral and have
\begin{eqnarray}
K_N(x,y)&=& c\,y^{N_A+N_B-N} \e^{-y/\sigma_A}\oint_{\gamma_{1}}\frac{dz_{1}}{2\pi i}
\oint_{\gamma_{2}}\frac{dz_{2}}{2\pi i}
\frac{\e^{z_{1}+z_{2}}}{z_{1}^{N_{A}+1}z_{2}^{N_{B}}}
\label{kernel3}\\
&&\times\sum_{j=1}^N
\,_{1}F_{1}\left(m+1-N_A;m+1;(\sigma_A^{-1}-\sigma_{Bj}^{-1})y
\right)\frac{1}{\sigma_{Bj}^{N_B-N+1}}\prod_{l\neq j}\frac{\left(x-z_{1}\,\sigma_{A}-z_{2}\,\sigma_{Bl}\right) }{(\sigma_{Bj}-\sigma_{Bl})}\,,
\nn
\end{eqnarray}
with the constant
\begin{equation}\label{c-def}
c=\frac{N_{A}!(N_{B}-N)!}{(N_A+N_B-N)!}\,.
\end{equation}
From eq.~(\ref{kernel3}) it is obvious that the kernel is expressible in terms of matrix invariants of $\Sigma_B$. Therefore we have to extend the product in eq.~(\ref{kernel3}) by the missing terms in $\sigma_{Bj}$. This can be achieved by introducing a contour integral in an auxiliary variable $z_3$ as
\begin{align}
K_N(x,y)=&c\, y^{N_A+N_B-N} \e^{-y/\sigma_A}\oint_{\gamma_{1}}\frac{dz_{1}}{2\pi i}
\oint_{\gamma_{2}}\frac{dz_{2}}{2\pi i}\oint_{\gamma_{3}}\frac{dz_{3}}{2\pi i}
\frac{\e^{z_{1}+z_{2}}}{z_{1}^{N_{A}+1}z_{2}^{N_{B}}}\frac{1}{\det[z_3\eins_N-\Sigma_B^{-1}]}\label{kernel4}\\
&\times\frac{\det[(x-z_1\sigma_A)\eins_N-z_2\Sigma_B]}{x-z_1\sigma_A-z_2z_3^{-1}}
\,_{1}F_{1}\left(m+1-N_A;m+1;(\sigma_A^{-1}-z_3)y
\right)\,,\nn
\end{align}
which replaces the sum. Here the contour integral over $\gamma_{3}$ has to be specified in the following way.
The contour $\gamma_3$ only encircles the positive eigenvalues of $\Sigma_B$, but excludes the pole $z_3=z_2/(x-z_1\sigma_A)$. Since $z_1$ and $z_2$ lie on the contours that encircle the origin and no other pole, we can choose the radii of these contours equal to $(x-\epsilon)/\sigma_A$ and $\epsilon\,{\min}_{j=1,\ldots,N}\left[\sigma_{Bj}^{-1}/2\right]$, respectively, with $x>\epsilon>0$. Then the contour $\gamma_3$ encircles the positive real axis starting from ${\min}_{j=1,\ldots,N}\left[\sigma_{Bj}^{-1}\right]$. In this way we have the desired realisation of the contours given in the representation~\eqref{kernel4}. Thereby the poles from the determinant $\det[z_3\eins_N-\Sigma_B^{-1}]$ yield the only contributions to the integral over $\gamma_3$.

The expression~\eqref{kernel4} is suitable for an asymptotic analysis at large matrix dimensions, as an alternative to the discussion of eq.~\eqref{kernel2} at the end of section \ref{sec:kern}. Here one has to deform the contours suitably such that the saddle-point analysis can be performed. We will not pursue this further as the aim of the present work is the derivation of exact results at finite matrix size.

\begin{figure}[!t]
\centerline{\includegraphics[width=0.47\textwidth]{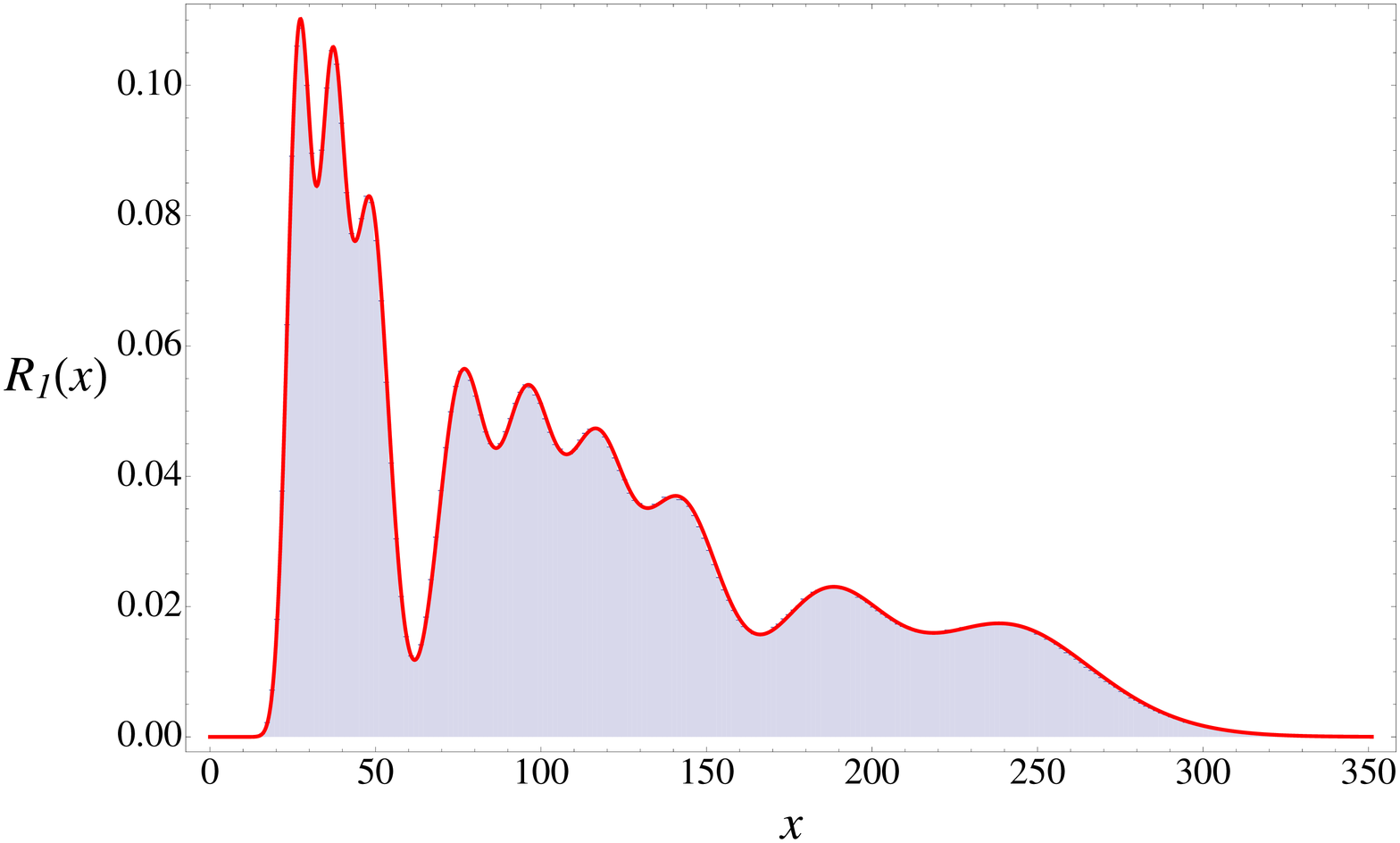}\hfill\includegraphics[width=0.47\textwidth]{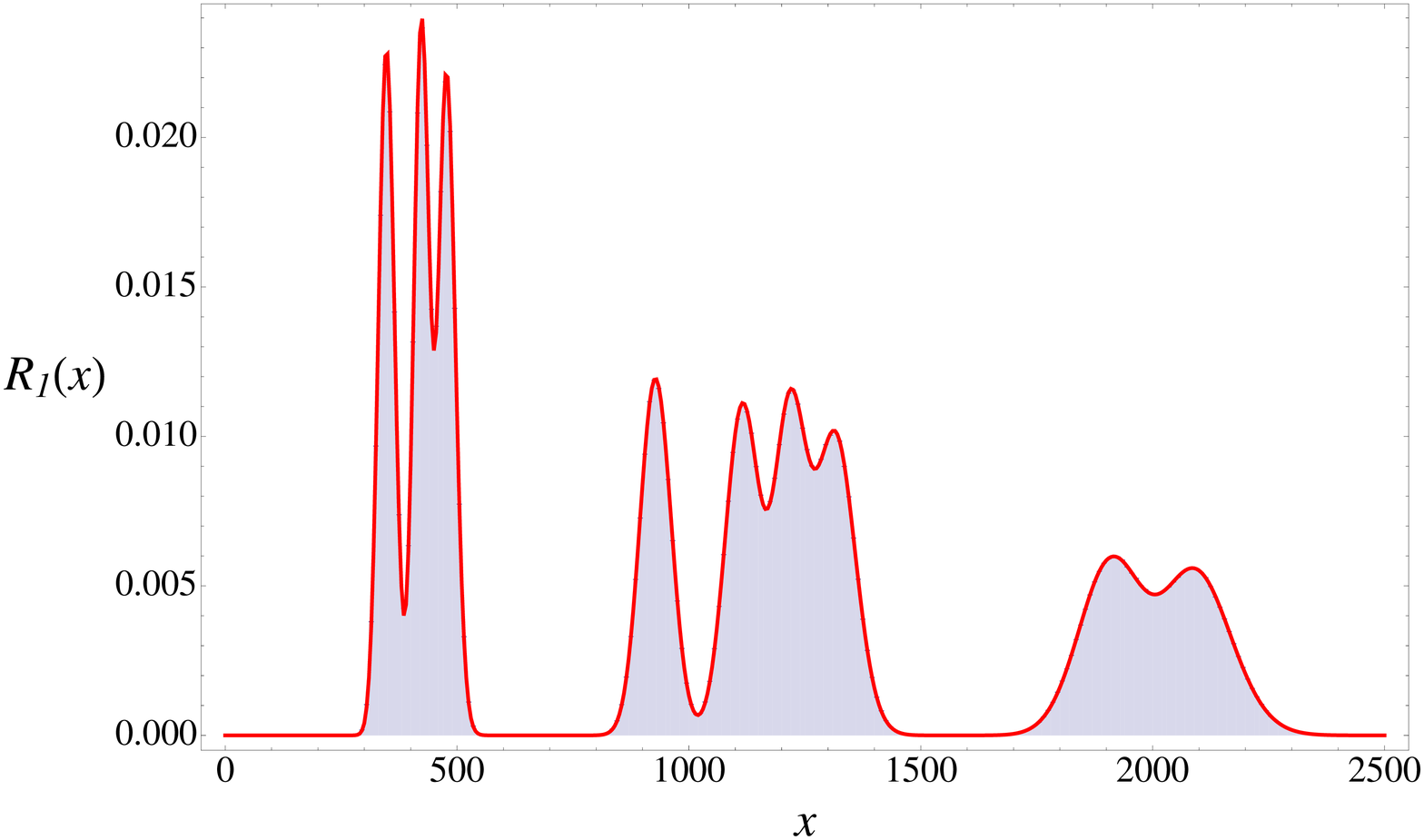}}
\caption{Comparison of the analytical result~\eqref{density} for the spectral density (red curves) with Monte Carlo simulations (histograms=shaded area, $10^6$ matrices drawn from the ensemble~\eqref{PABdef}). We employed the parameters $N=9$ with the fixed covariance matrices $\Sigma_A=\eins_9$ and $\Sigma_B=\diag(0.02, 0.20, 0.30, 1.50, 2.01, 2.25, 2.27, 4.05, 4.13)$ and the time length of the epochs $(N_A,N_B)=(35,40)$ (left plot) and $(N_A,N_B)=(350,400)$ (right plot). For longer epochs $N_A$ and $N_B$ the peaks overlap less compared to the shorter epochs. Hence they shift more and more towards the deterministic positions given by equation~\eqref{limit-largeNAB}. The remaining deviations from those positions result from the level repulsion amongst the individual eigenvalue distributions originating from the joint density (\ref{jpdf}) which are still visible.}
\label{fig}
\end{figure}

Following eq.~\eqref{kernel4} the spectral density is given by
\begin{eqnarray}
R_1(x)&=&  K_N(x,x)\nn\\
&=&c\, x^{N_A+N_B-N} \e^{-x/\sigma_A}\oint_{\gamma_{1}}\frac{dz_{1}}{2\pi i}
\oint_{\gamma_{2}}\frac{dz_{2}}{2\pi i}\oint_{\gamma_{3}}\frac{dz_{3}}{2\pi i}
\frac{\e^{z_{1}+z_{2}}}{z_{1}^{N_{A}+1}z_{2}^{N_{B}}}\frac{1}{\det[z_3\eins_N-\Sigma_B^{-1}]}\nn\\
&&\times\frac{\det[(x-z_1\sigma_A)\eins_N-z_2\Sigma_B]}{x-z_1\sigma_A-z_2z_3^{-1}}
\,_{1}F_{1}\left(m+1-N_A;m+1;(\sigma_A^{-1}-z_3)x
\right)\,.
\label{density}
\end{eqnarray}
Since all integrals can be evaluated by the residue theorem at the specific poles, the simplest way to compute this three-fold integral numerically is via its series expansion. In Fig.~\ref{fig} we show the spectral density for finite $N$ for two generically chosen sets of eigenvalues of $\Sigma_B$. Here we set $\sigma_A=1$, since $\sigma_{A}$ only rescales the spectrum. The density is compared to numerical simulations, with details given in Fig.~\ref{fig}, and we find an excellent agreement.

Let us try to understand some qualitative features of the density shown in Fig.~\ref{fig}. The position of the peaks of the spectrum can be easily estimated in the regime $N_A,N_B\gg N$. Indeed when scaling $N_A=n N'_A$, $N_B=n N'_B$ and $x=nx'$ with $N,N'_A,N'_B, x$ fixed, in the limit $n\to\infty$ we can perform a saddle-point approximation of the matrix representation of the ensemble in eq.~\eqref{PABdef}, yielding $H=n N'_A\Sigma_A+n N'_B\Sigma_B$. Then the limiting spectral density simplifies to
\begin{equation}\label{limit-largeNAB}
\lim_{n\to\infty} nR_1(nx')= \Tr \delta\left(x'\eins_N-N'_A\Sigma_A-N'_B\Sigma_B\right).
\end{equation}
This limit also holds in the general case where both matrices $\Sigma_A$ and $\Sigma_B$ are non-degenerate. At finite but large $N_A,N_B\gg N$ the peaks are broadened. When the distributions of the individual eigenvalues of $H$ overlap the eigenvalues repel each other and shift away from the deterministic positions~\eqref{limit-largeNAB}. This can be nicely seen in Fig.~\ref{fig}, where we considered two examples with $(N_A,N_B)=(35,40)$ and with $(N_A,N_B)=(350,400)$, keeping $N=9$ fixed. The sharpening of the peaks for larger values of $N_A$ and $N_B$ is obvious. Nonetheless the distributions of the individual eigenvalues still overlap. In a Gaussian approximation the spacing between neighbouring peaks scales as $N_A+N_B$ while the width of the peaks scales as $\sqrt{N_A+N_B}$. Thus a factor of ten in $N_A$ and $N_B$ yields a factor of about $1/\sqrt{10}\approx 1/3$, which explains the shape modification form the left to the right in Fig.~\ref{fig}.

\sect{Solution of the Non-Degenerate Case}\label{sec:nondeg}

In this section we will compute the generating function for the $k$-point density correlation function in the general case of non-degenerate $\Sigma_A\neq\Sigma_B$. 
In view of the distribution of matrix elements of $H$ being given by a hypergeometric function of matrix argument, see eq.~(\ref{Hjpdf}),  we have to choose a different strategy, due to the absence of any bi-orthogonal structure. The techniques we will use instead are supersymmetry and superbosonisation, generalising the computations from subsections \ref{sec:inv-single} and \ref{sec:exp-single}. 

Let us first redefine the $k$-point density correlation functions,
\be
\tilde{R}_k(\la_1,\ldots,\la_k)\equiv \left\langle \prod_{j=1}^k
\Tr \delta(\la_j\eins_{N}-H)  \right\rangle_{N,N_A,N_B}^{\Sigma_A,\Sigma_B}\, .
\label{Rktilde}
\ee
The matrix delta-functions in this expectation value can be generated by an appropriate differentiation of the following generating function for the $k$-point density correlation function,
\be
Z_{q|p}(X)\equiv \left\langle \frac{\prod_{j=1}^p\det[x_j\eins_{N}-H]}{\prod_{l=1}^q\det[y_l\eins_{N}-H]}
 \right\rangle_{N,N_A,N_B}^{\Sigma_A,\Sigma_B}\,,
\label{Zpqdef}
\ee
where we denote $X=\mbox{diag}(y_1,\ldots,y_q,x_1,\ldots,x_p)$, with $\im(y_l)\neq0$\, for all\, $l=1,\ldots,q$. The generating function $Z_{q|p}(X)$ defined in eq.~(\ref{Zpqdef}) is the central object of this section. 

As an important 
example let us consider
the spectral density 
$\tilde{R}_{1}(\lambda_1)$  with $k=1$.
In order to derive it let us define the averaged Green's function or resolvent as
\begin{align}
W_1(y)\equiv\left\langle \Tr\frac{1}{y\eins_{N}-H} \right\rangle_{N,N_A,N_B}^{\Sigma_A,\Sigma_B}\,.
\end{align}
It is well known that the spectral 
density can be obtained from the 
resolvent by taking its imaginary part:
\be\label{limit}
\tilde{R}_1(y)=\frac1\pi \lim_{{\rm Im}(y)\to 0^+} {\rm Im}(W_1(y))\,.
\ee
Note that the density $\tilde{R}_{1}(\lambda_1)$ is normalised to $N$. From eq.~(\ref{Zpqdef}) at $p=q=1=k$ the 
resolvent $W_{1}(y)$ results from the generating function 
by simple differentiation:
\be\label{Green-def}
\partial_{x}Z_{1|1}(X)\big|_{x=y}=W_1(y)\,.
\ee
In the same way the product of all $k$ Dirac delta-functions leading to the $k$-point density correlation functions $\tilde{R}_k(\la_1,\ldots,\la_k)$ in eq.~(\ref{Rktilde}) can be generated from $Z_{k|k}(X)$ by successive differentiation 
leading to the $k$-point resolvent, 
and by taking imaginary parts to zero from above as in eq.~(\ref{limit}). 

It is well known \cite{Guhr:1998} that for $k>1$ the $k$-point density correlation functions defined in eq.~(\ref{Rktilde}) differ from those defined in eq.~(\ref{Rkdef1}) by so-called contact terms, e.g. for $k=2$
\be
\tilde{R}_2(\la_1,\la_2) = R_2(\la_1,\la_2) + \delta(\la_1-\la_2) R_1(\la_1)\ ,
\ee
where the last term originates from coinciding arguments of $k=2$ Dirac delta-functions in  eq.~(\ref{Rktilde}).

We would like to add that for $q=0$ in eq.~(\ref{Zpqdef}) the quantity $Z_{0|p}$ can be used in combination with the replica method to compute
the generating function of the ergodic capacity, the average of $\ln[\det[\eins+H]]$. As mentioned in the introduction this quantity is of central interest in applications to telecommunications in the setup of MIMO.

The strategy to compute all ratios of characteristic polynomials in the generalised partition function $Z_{q|p}(X)$ will be similar to the computation in subsection~\ref{sec:exp-single}. We rewrite the determinants as Gaussian integrals, now over commuting and anti-commuting variables, then integrate over $W$, apply the duality 
between ordinary and supermatrix spaces
as in eq.~(\ref{Qdef}) and in the end employ the corresponding superbosonisation formula~\cite{BEKYZ,Som07,BA07,LSZ08,KSG09}. The details of this computation are carried out in appendix~\ref{susyiso}. The final result for $q\leq N$ is
\begin{eqnarray}\label{gen-res}
Z_{q|p}(X)&=&C_{N_A}C_{N_B}\int d\mu(U_A) \int d\mu(U_B)\e^{-\Str U_A-\Str U_B} \Sdet^{N_A}\left(U_A\right)\,\Sdet^{N_B}\left(U_B\right)\\
&&\times\Sdet^{-1}(\eins_N\otimes X-\Sigma_A\otimes U_A-\Sigma_B\otimes U_B)\,,\nonumber
\end{eqnarray}
where the constants are given by
\begin{equation}\label{constant}
C_n=\prod_{l=0}^{q-1}\frac{1}{\pi^{l}(n-q+l)!}\prod_{l=0}^{p-1}\frac{(n-q+l)!}{\pi^{l}}\,.
\end{equation}
The definition of the supermatrices $U_{A},U_B\in{\rm Herm}_+(q|p)$, the supertrace $\Str(\ldots)$ and the superdeterminant $\Sdet(\ldots)$ are recalled in appendix~\ref{sec:recall}. The Haar measure is given in eq.~\eqref{Haar-susy}. 

As a check for $\Sigma_A=\Sigma_B=\Sigma$, eq.~(\ref{gen-res}) simplifies to the well-known supersymmetric result~\cite{RKG10,RKGZ12,KKG14} of a single correlated Wishart-Laguerre ensemble,
\begin{eqnarray}\label{gen-res-simpl}
Z_{q|p}(X)|_{\Sigma_A,\Sigma_B=\Sigma}&=&C_{N_A+N_B}\int d\mu(U) \e^{-\Str U} \Sdet^{N_A+N_B}\left(U\right)\,\Sdet^{-1}(\eins_N\otimes X-\Sigma\otimes U)\,.\quad
\end{eqnarray}
This result can be readily deduced from eq.~\eqref{gen-res} via the two substitutions $U_B=U_A^{1/2}U'U_A^{1/2}$ and $U_A=(\eins_N+U')^{-1/2}U(\eins_N+U')^{-1/2}$. Here we have used the group invariance of the Haar measure.

In the case of the spectral density we need the case $(q|p)=(1|1)$. Then the Haar measure is simply $d\mu(U)=(2\pi i)^{-1} [dU]$, with $[dU]$ the flat measure. After a brute force expansion in the Grassmann variables and the integration over the two angles the generating function reduces to the one shown in eq.~\eqref{calc-app2}. For the calculation we refer to appendix~\ref{sec:density-calc}. The derivative with respect to $x$ at $x=y$ yields the spectral density,
\begin{align}
&\tilde R_1(y)=\frac{1}{\pi(N_A-1)!(N_B-1)!}\label{level-result}\\
&\times\lim_{{\rm Im}\,y\to0^+}{\rm Im}\left(1+\frac{\partial}{\partial\mu_A}\right)^{N_A-1}\left(1+\frac{\partial}{\partial\mu_B}\right)^{N_B-1}\int_0^\infty ds_A\int_0^\infty ds_B \e^{-s_A-s_B} s_A^{N_A}s_B^{N_B}\det [F_s]\nonumber\\
&\times\biggl(\frac{1}{s_As_B}\left(1+\frac{\partial}{\partial\mu_A}\right)\left(1+\frac{\partial}{\partial\mu_B}\right)+\frac{1}{s_B}\left(1+\frac{\partial}{\partial\mu_B}\right)\frac{\partial^2}{\partial z_A\partial z_A^*}\nonumber\\
&+\frac{1}{s_A}\left(1+\frac{\partial}{\partial\mu_A}\right)\frac{\partial^2}{\partial z_B\partial z_B^*} +\frac{\partial^4}{\partial z_A\partial z_A^*\partial z_B\partial z_B^*}\biggl)\nonumber\\
&\times\det\left[y\eins_N-\mu_A\Sigma_A-\mu_B\Sigma_B+\left[\begin{array}{cc} z_A\Sigma_A, & z_B\Sigma_B \end{array}\right]F_s\otimes\left[\begin{array}{cc} 1 & i \\ i & 1 \end{array}\right]\left[\begin{array}{c} z_A^*\Sigma_A \\ z_B^*\Sigma_B \end{array}\right]\right]\nonumber\\
&\times \left.\Tr\left(y\eins_N-\mu_A\Sigma_A-\mu_B\Sigma_B+\left[\begin{array}{cc} z_A\Sigma_A, & z_B\Sigma_B \end{array}\right]F_s\otimes\left[\begin{array}{cc} 1 & i \\ i & 1 \end{array}\right]\left[\begin{array}{c} z_A^*\Sigma_A \\ z_B^*\Sigma_B \end{array}\right]\right)^{-1}\right|_{\mu_{A/B}=z_{A/B}=0},\nonumber
\end{align}
with
\begin{equation}\label{abbreviation}
 F_s=(y\eins_N-s_A\Sigma_A-s_B\Sigma_B)^{-1}.
\end{equation}
The derivatives in $\mu_{A/B}$ and $z_{A/B}$ encode the integrals we carried out and can be easily numerically evaluated via a series expansion. The two-fold integral over $s_A$ and $s_B$ is the numerically non-trivial part, since it does not factorise. The limit of vanishing imaginary part, 
${{\rm Im}(y)\to0^+}$, reduces one of these remaining integrals to a Dirac delta-function. Thus we effectively end up with a sum of one-dimensional integrals, which can be evaluated numerically. A further and a more efficient way to evaluate the latter expression numerically is to adjust the imaginary increment. It can be chosen small enough for a result independent of the imaginary increment, while it is big enough for a numerically stable evaluation.

The expression for the spectral density simplifies a lot in the case when $N_A$, $N_B$ and $N$ become large. We will not perform a rigorous asymptotic analysis here but give a sketch of the expected answer for the limiting global or macroscopic density. Our approach complements the ideas for taking the large-$N$ limit sketched in section~\ref{sec:sol-hel-deg}. For this purpose we follow the ideas of refs.~\cite{WKG,WWKK} and start with the 
resolvent
in the supersymmetric formulation
\begin{eqnarray}\label{Green-SUSY}
W_1(y)&=&\frac{1}{ Z_{1|1}(y\eins_{1|1})}\int d\mu(U_A) \int d\mu(U_B)\frac{\e^{-\Str U_A-\Str U_B} \Sdet^{N_A}\left(U_A\right)\,\Sdet^{N_B}\left(U_B\right)}{\Sdet(y\eins_N\otimes \eins_{1|1}-\Sigma_A\otimes U_A-\Sigma_B\otimes U_B)}\\
&&\times\Str\left[(y\eins_N\otimes \eins_{1|1}-\Sigma_A\otimes U_A-\Sigma_B\otimes U_B)^{-1}\left(\begin{array}{cc}  \eins_N &0 \\ 0 & 0 \end{array}\right)\right]\,.\nonumber
\end{eqnarray}
The normalisation can be indeed chosen as $1/Z_{1|1}(y\eins_{1|1})$ which is certainly equal to $1$, with the advantage that we see which terms in the saddle-point approximation cancel. The saddle-point approximation starts with the Lagrangian
\begin{equation}\label{Lagrangian}
 \mathcal{L}(U_A,U_B)=\Str\, (U_A+U_B)-N_A\Str\,{\rm ln}\, U_A-N_B\Str\,{\rm ln}\, U_B+\Str\,{\rm ln}(y \eins_N\otimes\eins_{1|1}-\Sigma_A\otimes U_A-\Sigma_B\otimes U_B),
\end{equation}
where all superdeterminants have been raised to the exponent. From the first derivatives of the Lagrangian the saddle-point conditions, $\partial_{U_A}\mathcal{L}(U_A^{(0)},U_B^{(0)})=0$ and $\partial_{U_B}\mathcal{L}(U_A^{(0)},U_B^{(0)})=0$, yield
\begin{equation}\label{Saddle-eq}
\eins_{1|1}-N_A{U_A^{(0)}}^{-1}-\Tr_N\left[(y \eins_N\otimes\eins_{1|1}-\Sigma_A\otimes U_A^{(0)}-\Sigma_B\otimes U_B^{(0)})^{-1}\Sigma_A\right]=0\ ,
\end{equation}
and the second equation results from interchanging $A$ and $B$. The notation $\Tr_N$ indicates the partial trace in the ordinary $N\times N$ matrix space, such that the result of this trace operation is an $(1|1)\times (1|1)$ supermatrix. The remaining trace over this supermatrix is denoted by $\Str_{(1|1)}$. Assuming the uniqueness of the contributing saddle point $(U_A^{(0)},U_B^{(0)})$, we have for the spectral density
\begin{eqnarray}\label{level-asymp}
\tilde R_1(y)&=&\frac{1}{\pi}
{\rm Im}\,\Str\left[(y\eins_N\otimes \eins_{1|1}-\Sigma_A\otimes U_A^{(0)}-\Sigma_B\otimes U_B^{(0)})^{-1}\left(\begin{array}{cc} 
\eins_N  &0 \\ 0 & 0 \end{array}\right)\right]\nonumber\\
&=&\frac{1}{\pi}
{\rm Im}\,\Str_{(1|1)}\left[ 
\Tr_N\left[(y\eins_N\otimes \eins_{1|1}-\Sigma_A\otimes U_A^{(0)}-\Sigma_B\otimes U_B^{(0)})^{-1}\right]
\left(\begin{array}{cc} 
1  &0 \\ 0 & 0 \end{array}\right)\right].
\end{eqnarray}
Note that we have already taken the limit ${\rm Im}(y)\to0^+$ here which is possible due to the regularity of the saddle-point solution. To see that no Grassmann variables are involved in this expression (otherwise it would be inconsistent) we multiply eq.~\eqref{Saddle-eq} with $U_A^{(0)}$ and add it to the corresponding equation for $U_B^{(0)}$. This leads to
\begin{equation}\label{Saddle-eq-new}
U_A^{(0)}+U_B^{(0)}-(N_A+N_B-N)\eins_{1|1}-y\Tr_N\left[(y \eins_N\otimes\eins_{1|1}-\Sigma_A\otimes U_A^{(0)}-\Sigma_B\otimes U_B^{(0)})^{-1}\right]=0.
\end{equation}
We can insert this equation into eq.~\eqref{level-asymp} and arrive at the simpler expression
\begin{equation}\label{level-asymp.b}
\tilde R_1(y)=\frac{1}{\pi}{\rm Im}\,\Str_{(1|1)}\left[\frac{1}{y}(U_A^{(0)}+U_B^{(0)})\left(\begin{array}{cc}1 &0 \\ 0 &0 \end{array}\right)\right].
\end{equation}
Hence only the diagonal elements of the saddle point $(U_A^{(0)},U_B^{(0)})$ are involved.

Analysing the saddle-point equation~\eqref{Saddle-eq} in more detail requires some effort. However, as we have already seen in the exact expression~\eqref{level-result} for the spectral density, the Grassmann variables only contribute to the integrand as a polynomial of a very small order. Therefore they do not enter the saddle-point analysis. Thanks to the projection matrix in eq. (\ref{level-asymp.b}) only the upper boson-boson entries of the matrices  $U_A^{(0)}$ and  $U_B^{(0)}$ enter, denoted by $q_A^{(0)}$ and $q_B^{(0)}$, respectively. They are determined by the two coupled saddle-point equations for these entries, eq. (\ref{Saddle-eq}) and its counterpart:
\begin{eqnarray}\label{Saddle-eq-b}
1-\frac{N_A}{q_A^{(0)}}-\Tr\left[(y \eins_N-q_A^{(0)}\Sigma_A -q_B^{(0)}\Sigma_B)^{-1}\Sigma_A\right]&=&0,\\
1-\frac{N_B}{q_B^{(0)}}-\Tr\left[(y \eins_N-q_A^{(0)}\Sigma_A -q_B^{(0)}\Sigma_B)^{-1}\Sigma_B\right]&=&0\ .\nonumber
\end{eqnarray}
The two unknown complex functions $q_A^{(0)}(y)$ and $q_B^{(0)}(y)$ depend on the variable $y$. The asymptotic of the spectral density thus simplifies to the final result
\begin{equation}\label{level-asymp.c}
\tilde R_1(y)=\frac{1}{\pi}
{\rm Im}\left(\frac{1}{y}(q_A^{(0)}(y)+q_B^{(0)}(y))\right).
\end{equation}
This expression is reminiscent to the one presented in ref.~\cite{WKG} for the spectral density of a single correlated Wishart matrix. However, the evaluation of the corresponding saddle-point equation in~\cite{WKG} compared to those in eq.~(\ref{Saddle-eq-b}) was much simpler because of two points.

First, one is looking for only one function $q(y)$ in the case of a single correlated Wishart matrix, satisfying a single and not two coupled equations.
In our case there are $(N+1)^2$ solutions including multiplicity due to B\'ezout's theorem, underlining the complexity of the problem. These solutions have to come either in complex conjugate pairs or are real because we can complex conjugate both equations.
Due to the imaginary increment of $y$ only one half of the solutions can be reached with the integration of $s_A$ and $s_B$, see the exact result~\eqref{level-result}.

Second, only one empirical correlation matrix $\Sigma$ is involved for a single Wishart matrix~\cite{WKG}. Thus we can diagonalise this matrix, which is not possible in the most general case for the present random matrix model with two empirical correlation matrices, $\Sigma_A\neq \Sigma_B$ satisfying $[\Sigma_A,\Sigma_B]\neq0$. Nonetheless, the range of applicability of the strikingly simple result~\eqref{level-asymp.c} should be as good as for the simpler model studied in ref.~\cite{WKG}. We close the discussion at this point because the main goal of the present work is the exact finite-$N$ solution and not the detailed asymptotic analysis.

Let us make a few final remarks on generalisations of the result~\eqref{gen-res}. One can easily extend this to real ($\beta=1$) and quaternion ($\beta=4$) matrices $H$. The symmetries of the supermatrices change accordingly, where the positive definite Hermitian boson-boson blocks of $U_A$ and $U_B$ become either symmetric or self-dual and the unitary fermion-fermion blocks become self-dual or symmetric, respectively, see \cite{LSZ08,KSG09,KKG14,Kieburg15}. Moreover, these supermatrices become twice as large for both $\beta=1,4$. Another point, which only plays a role in the real case, is the exponent of the superdeterminants~\eqref{gen-res}. For $\beta=1$ one has to take the square roots of all three superdeterminants, while for $\beta=4$ the exponents remain the same.

Moreover, with the focus on a generalisation of the generating function~(\ref{gen-res}) one can replace the Gaussian weights for $A$ and $B$ in eq.~(\ref{ProbDist}) by other weights from invariant ensembles like the Jacobi or the Cauchy-Lorentz ensemble cf. \cite{Levy-tails,Nowak,product} for recent related works. A simple way to calculate the according expression is shown in the works~\cite{KKG14,Kieburg15} and is called projection formula. It is a shortcut of the map to superspace and does not only show that the weights in superspace (here the terms $\mathcal{Q}_A(U_A)\mathcal{Q}_B(U_B)\propto\exp[-\Str U_A-\Str U_B]$) have to be replaced, but also what these weights $\mathcal{Q}_A$ and $\mathcal{Q}_B$ are as functionals of the probability distributions $\mathcal{P}_A$ and $\mathcal{P}_B$ in ordinary space.

For another generalisation, more than two epochs can be introduced, namely the matrix $H=A_1A_1^\dagger+\ldots+A_TA_T^\dagger$ with correlation matrices $\Sigma_{A_1},\ldots,\Sigma_{A_T}$ and weights $\mathcal{P}_{A_1},\ldots,\mathcal{P}_{A_T}$ for arbitrary $T$ can be considered. Then the result~\eqref{gen-res} can 
be extended to the product of $T$ integrals,
\begin{align}\label{gen-res-gen}
Z_{q|p}(X)=&\int\prod_{j=1}^T C_{N_{A_j}}d\mu(U_{A_j}) \mathcal{Q}_{A_j}(U_{A_j})\Sdet^{N_{A_j}}\left(U_{A_j}\right)\\
&\times \Sdet^{-1}\left(\eins_N\otimes X-\sum_{k=1}^T\Sigma_{A_k}\otimes U_{A_k}\right).\nonumber
\end{align}
As mentioned above, one could also choose weights $\mathcal{P}_{A_j}$ different from eq. (\ref{PABdef}), e.g. by choosing  Gaussian ensembles of Hermitian matrices or ensembles of Jacobian or Cauchy-Lorentzian type. Then the first line of~(\ref{gen-res-gen}) changes accordingly in the supersymmetric weights $\mathcal{Q}_{A_j}$, depending on the initial weights $\mathcal{P}_{A_j}$. In the second line of eq.~(\ref{gen-res-gen}) the substructure of the generalisation to a  matrix $H$ as the sum of $T$ correlation matrices is clear. It is astonishing that our result for the generating function for $H$ as a sum of $T=2$ Wishart correlation matrices~\eqref{gen-res} 
enjoys a generalisation to much more complicated ensembles.

\sect{Discussion and Outlook}\label{sec:conclusio}

In the present work  we determined the spectral statistics of all $k$-point eigenvalue density correlation functions for the sum $H$ of two correlated Wishart matrices. In previous work by Kumar it was shown that the joint density of the eigenvalues of $H$ is given by a hypergeometric function of matrix argument. It was also shown in his work that in the half-degenerate case, when one of the correlation matrices is proportional to the identity, the joint density becomes a bi-orthogonal ensemble. It is given by the Laguerre weight times a Vandermonde determinant and a determinant of ordinary confluent hypergeometric functions of Kummer type $_1F_1$. We first solved this half-degenerate case by determining its kernel of bi-orthogonal functions explicitly, exploiting the bi-orthogonal structure. Remarkably, the kernel could be constructed by computing the expectation value of a single characteristic polynomial, providing the polynomials orthogonal to $_1F_1$. This expectation value was computed using supersymmetry and bosonisation in the general non-degenerate case. As a byproduct we computed the expectation value of the inverse of a single characteristic polynomial in this general case too, as well as a more compact form of the normalisation constant of the joint density in the half-degenerate case. The solution for the $k$-point correlation function is then given in terms of a  determinant of size $k$ of this kernel. In contrast the previous representation derived by Kumar was given in terms of a determinant of size $N+k$, containing the moments of $_1F_1$. Our result thus offers the possibility to take the large-$N$ limit and to study questions of universality by making an asymptotic expansion of the kernel. In addition to the sum over bi-orthogonal functions we gave an alternative three-fold complex contour integral representation of our kernel. Both representations are amenable to take the large-$N$ limit. We illustrated our finite-$N$ result for the spectral density with Monte-Carlo simulations.

In the more difficult case where both correlation matrices are non-degenerate no bi-orthogonal structure is available. Therefore we applied the standard supersymmetric method and computed the generating functions for the $k$-point resolvents. They are given by the expectation values of $k$ ratios of characteristic polynomials that we computed using the superbosonisation formula. They are given by two integrals over symmetric supermanifolds of $k$-dependent dimension. As an example we spelled out the generating function for the spectral density by choosing an explicit parametrisation for $k=1$. It can be written as an integral over two real positive variables and two angles, containing determinants and traces of rational functions of the correlation matrices and the four integration variables. The angular integrals could be solved, leading to derivatives of the order of the matrix dimensions. Based on these results we derived two coupled saddle-point equations in two variables in the large-$N$ limit and expressed the limiting spectral density in terms of their solution.

Several open questions are left for future work. This includes in particular the asymptotic analysis of the limiting kernel from the first part of our investigations, but also a detailed analysis of the saddle-point equations for the density in the more general non-degenerate case from the second part of our work. In this second part we also gave indications how our results could be generalised in several directions. This includes different symmetry classes with Wishart matrices built from real or quaternionic matrix elements. We also wrote down the general structure of the generating function when the matrix $H$ consists of the sum of $T\geq2$ correlated Wishart matrices in the non-degenerate case. This structure persists even when the Gaussian weights of the $T\geq2$ correlated Wishart matrices are replaced by more general weights of ensembles of Jacobi or Cauchy-Lorentzian type. Of course in this case the corresponding supersymmetric weights inside the integrals over the supermanifolds have to be replaced by their Jacobi or Cauchy-Lorentzian counterparts.

\section*{Acknowledgements}

We thank Santosh Kumar for fruitful discussions. Financial support through LabEx PALM grant number ANR-10-LABX-0039-PALM (G.A.), FSPM$^2$ (T.C.) and CRC 701: \textit{Spectral Structures and Topological Methods in Mathematics} of the German Research Council DFG (M.K.) are gratefully acknowledged. Two of us (G.A. and T.C.) would like to kindly thank the LPTMS in Orsay for hospitality where part of this work was established.

\begin{appendix}

\sect{Simplification of the Joint Probability Density}\label{sec:alter}

In this appendix we present an alternative derivation of the joint probability density $P_N$, see eq.~(\ref{jpdf}). The benefit of this computation is twofold. First, we obtain a more explicit form of the normalisation constant in eq.~(\ref{jpdf-const}) given by a product of Gamma functions and the eigenvalues of the non-trivial covariance matrix. This has to be compared to the determinant of a Gauss type hypergeometric function in eq.~(\ref{norm}). Second, we show that in the special case of parameters at hand the confluent hypergeometric functions inside the determinant of the joint probability density (\ref{jpdf}) can be expressed in terms of more elementary functions.

To calculate the joint probability density of the matrix $H=AA^\dagger +BB^\dagger$ we introduce a test function $f(H)$, which is a Schwartz function on the space ${\rm Herm}(N)$ of Hermitian $N\times N$ matrices and satisfies the relation
\begin{equation}\label{app-inv}
 f(H)=f(UHU^\dagger),\ {\rm for\ all}\ H\in{\rm Herm}(N)\ \mbox{and}\ U\in{\rm U}(N)\,.
\end{equation}
Then for any function $f$ with these properties we have that
\begin{align}
\langle f(AA^\dagger+BB^\dagger)\rangle_{N,N_A,N_B}^{\Sigma_A,\Sigma_B}=\int 
[dH]\ {\cal P}_H(H) f(H)=\int[d\Lambda]\ {\cal P}_N(\lambda_1,\ldots,\lambda_N)f(\Lambda)\,.\label{app-jpdf-def}
\end{align}
In the sense of weak topology this is the definition of the joint probability density ${\cal P}_N$ of eigenvalues $\Lambda={\rm diag}(\lambda_1,\ldots,\lambda_N)$ of the combined random matrix $H=AA^\dagger +BB^\dagger$.
We make use of exactly this definition to derive an alternative expression for the joint probability density $P_N$ in the half-degenerate case which is more explicit than the one derived in~\cite{Kumar:2014}.

In a first step in deriving the joint probability density we introduce the Dirac delta-function
\begin{align}
\delta(H-AA^\dagger-BB^\dagger)=&\prod_{j=1}^N\delta(H_{jj}-(AA^\dagger+BB^\dagger)_{jj})\label{app-delta}\\
&\times\prod_{1\leq i<j\leq N}\delta(\re[H_{ij}-(AA^\dagger+BB^\dagger)_{ij}])\delta(\im[H_{ij}-(AA^\dagger+BB^\dagger)_{ij}])\,.\nonumber
\end{align}
Those Dirac delta-functions can be written as a double Fourier transform yielding
\begin{eqnarray}\label{a-der-1}
&&\langle f(AA^\dagger+BB^\dagger)\rangle_{N,N_A,N_B}^{\Sigma_A,\Sigma_B}\\
&=&\frac{{\det}^{-N_A}\Sigma_A{\det}^{-N_B}\Sigma_B}{\pi^{N(N_A+N_B)}} \frac{1}{2^N\pi^{N^2}}\lim_{t\to0}\int [dA] \exp[-\Tr\Sigma_A^{-1}AA^\dagger]\int [dB]\exp[-\Tr\Sigma_B^{-1}BB^\dagger] \nonumber\\
&&\times\int [dH] f(H)\int [dK] \exp[-t\Tr K^2] \exp[i \Tr K(H-AA^\dagger-BB^\dagger)]\,.\nonumber
\end{eqnarray}
We also introduced a regularising Gaussian factor $\exp[-t\Tr K^2]$ with auxiliary parameter $t$, which we send to zero in the end. In this way all four sets of integrals over $A$, $B$, $H$ and $K$ are absolutely integrable and, thus, can be interchanged. The integral over $H$ is absolutely integrable because of the test function $f$, which is also the reason for employing this kind of function.

The first fraction in the constant in front of the integral~\eqref{a-der-1} is the one of the original probability weights~\eqref{PABdef}. The second fraction is the normalisation of the double Fourier transform such that the Dirac delta-functions are normalised to unity.

In a second step we integrate over the matrices $A$ and $B$. Hence eq.~\eqref{a-der-1} becomes
\begin{eqnarray}
\langle f(AA^\dagger+BB^\dagger)\rangle_{N,N_A,N_B}^{\Sigma_A,\Sigma_B}
&=&\frac{{\det}^{-N_A}\Sigma_A{\det}^{-N_B}\Sigma_B}{2^N\pi^{N^2}}\lim_{t\to0} \int [dH] f(H)\int [dK] \exp\left[-t\Tr K^2+i \Tr KH\right]\nonumber\\
&&\times{\det}^{-N_A}\left[\Sigma_A^{-1}+i K\right]{\det}^{-N_B}\left[\Sigma_B^{-1}+i K\right].\label{a-der-2}
\end{eqnarray}
To proceed further we have to simplify the ensemble since we want to diagonalise the matrices $H=V\Lambda V^\dagger$ and $K=U\Omega U^\dagger$ and we need to integrate over the cosets $V,U\in{\rm U}(N)/[{\rm U}^N(1)\times\mathbb{S}(N)]$. The set $\mathbb{S}(N)$ is the permutation group of $N$ elements and lifts the ordering of the eigenvalues $\lambda_{j}$ and $\omega_{j}$ of $H$ and $K$, respectively.

At this point of the calculation we assume $\Sigma_A=\sigma_A\eins_N$. Then we can also assume that $\Sigma_B={\rm diag}(\sigma_{B1},\ldots,\sigma_{BN})$ because the whole system is invariant under adjoint transformations $\Sigma_B\to O\Sigma_B O^\dagger$ for all unitary matrices $O\in{\rm U}(N)$. Diagonalising $H$ and $K$ and using the invariance of the normalised Haar measure of the coset ${\rm U}(N)/[{\rm U}^N(1)\times\mathbb{S}(N)]$ under $V\to UV$ we have
\begin{eqnarray}\label{a-der-3}
&&\langle f(AA^\dagger+BB^\dagger)\rangle_{N,N_A,N_B}^{\Sigma_A,\Sigma_B}\\
&=&\frac{{\det}^{-N_A}\Sigma_A{\det}^{-N_B}\Sigma_B}{(2\pi)^N\prod_{l=1}^N[l!]^2}\lim_{t\to0} \int [d\Lambda] \Delta_N^2(\{\lambda_j\})f\left(\Lambda\right)\int [d\Omega] \Delta_N^2(\{\omega_j\})\e^{-t\Tr \Omega^2}{\det}^{-N_A}\left[\sigma_A^{-1}\eins_N+i \Omega\right]\nonumber\\
&&\times\left(\int_{-\infty}^\infty d\mu(U)\exp[i\Tr U\Omega U^\dagger\Lambda]\right)\left(\int d\mu(V){\det}^{-N_B}\left[\sigma_B^{-1}+i V\Omega V^\dagger\right]\right).\nonumber
\end{eqnarray}
The additional constant is the squared volume of the coset ${\rm U}(N)/[{\rm U}^N(1)\times\mathbb{S}(N)]$. The square is needed since we have one coset for $U$ and one for $V$ and the Haar measures are normalised.

The two coset integrals are well-known. The first integral is the Harish-Chandra-Itzykson-Zuber integral~\cite{Harish,ItzZub}
\begin{equation}\label{HCIZ}
 \int d\mu(U)\exp[i\Tr U\Omega U^\dagger\Lambda]=\left(\prod_{l=0}^{N-1}\frac{l!}{i^l}\right)\frac{\det\left[\exp[i \omega_a\lambda_b]|_{1\leq a,b\leq N}\right]}{\Delta_N(\{\lambda_j\})\Delta_N(\{\omega_j\})}\,,
\end{equation}
where the normalisation is fixed to unity at $\Omega=0$. The second integral is also known~\cite{HO06}
\begin{equation}\label{det-int}
 \int d\mu(V){\det}^{-N_B}\left[\sigma_B^{-1}+i V\Omega V^\dagger\right]=\left(\prod_{l=0}^{N-1}\frac{ l!(N_B-N)!}{i^l(N_B-N+l)!}\right)\frac{\det\left[(\sigma_{B,b}^{-1}+i\omega_a)^{N-N_B-1}|_{1\leq a,b\leq N}\right]}{\Delta_N(\{\omega_j\})\Delta_N(\{\sigma_{Bj}^{-1}\})}\,.
\end{equation}
Again the normalisation can be fixed by taking $\Omega=0$ which yields ${\det}^{-N_B}\Sigma_B$. We insert these two integrals into eq.~\eqref{a-der-3} and find
\begin{eqnarray}\label{a-der-4}
&&\langle f(AA^\dagger+BB^\dagger)\rangle_{N,N_A,N_B}^{\Sigma_A,\Sigma_B}\\
&=&\frac{{\det}^{-N_A}\Sigma_A{\det}^{N-N_B-1}\Sigma_B}{(2\pi)^N[N!]^2\Delta_N(\{\sigma_{Bj}\})}\left(\prod_{l=0}^{N-1}\frac{(N_B-N)!}{(N_B-N+l)!}\right)\lim_{t\to0} \int [d\Lambda] \Delta_N(\{\lambda_j\})f(\Lambda)\nonumber\\
&&\times\int [d\Omega] \e^{-t\Tr \Omega^2}{\det}^{-N_A}\left[\sigma_A^{-1}\eins_N+i \Omega\right]\det\left[\e^{i \omega_a\lambda_b}|_{1\leq a,b\leq N}\right]\det\left[(\sigma_{B,b}^{-1}+i\omega_a)^{N-N_B-1}|_{1\leq a,b\leq N}\right].\nonumber
\end{eqnarray}

To evaluate the integral over $\Omega$ we apply Andr\'eief's integration formula~\cite{Andr}, see eq.~\eqref{genAnd} for $k=l=0$. Thereby we notice that the integral over $\Omega$ is absolutely integrable even at $t=0$ because of $N_A,N_B\geq N$, such that we can set $t=0$ from now on. Thus we end up with
\begin{eqnarray}\label{a-der-5}
&&\langle f(AA^\dagger+BB^\dagger)\rangle_{N,N_A,N_B}^{\Sigma_A,\Sigma_B}\\
&=&\frac{{\det}^{-N_A}\Sigma_A{\det}^{N-N_B-1}\Sigma_B}{N!\Delta_N(\{\sigma_{Bj}\})}\left(\prod_{l=0}^{N-1}\frac{(N_B-N)!}{(N_B-N+l)!(N_A+N_B-N)!}\right)\int [d\Lambda] \Delta_N(\{\lambda_j\})f(\Lambda)\nonumber\\
&&\times\det\left[\left.(N_A+N_B-N)!\int \frac{d\kappa}{2\pi} \frac{\exp[i \kappa\lambda_a]}{(\sigma_A^{-1}+i \kappa)^{N_A}(\sigma_{B,b}^{-1}+i\kappa)^{N_B-N+1}}\right|_{1\leq a,b\leq N}\right].\nonumber
\end{eqnarray}
Now we are ready to identify the explicit form of the normalisation constant
\begin{equation}\label{const-jpdf} C_{N,N_A,N_B}^{\Sigma_A,\Sigma_{B}}=\frac{{\det}^{-N_A}\Sigma_A{\det}^{N-N_B-1}\Sigma_B}{N!\Delta_N(\{\sigma_{Bj}\})}\left(\prod_{l=0}^{N-1}\frac{(N_B-N)!}{(N_B-N+l)!(N_A+N_B-N)!}\right),
\end{equation}
of the joint probability density $P_N(\lambda_1,\ldots,\lambda_N)$ and the functions inside the determinant in eq. (\ref{a-der-5})
\begin{equation}\label{phi-jpdf-1}
 \varphi_j(\lambda)=(N_A+N_B-N)!\int \frac{d\kappa}{2\pi} \frac{\exp[i \kappa\lambda]}{(\sigma_A^{-1}+i \kappa)^{N_A}(\sigma_{Bj}^{-1}+i\kappa)^{N_B-N+1}}\,.
\end{equation}
These functions are normalised such that $\lim_{\lambda\to0}\lambda^{N-N_A-N_B}\varphi_j(\lambda)=1$. They can be calculated via the residue theorem. Thereby we only get a contribution when $\lambda$ is positive. Then we can close the contour around the two poles $i\sigma_A^{-1}$ and $i\sigma_{Bj}^{-1}$. This yields two contributions, and we have explicitly
\begin{eqnarray}\label{phi-jpdf-2} \varphi_j(\lambda)&=&\frac{(N_A+N_B-N)!}{(N_A-1)!}\left(-\frac{\partial}{\partial \sigma_A^{-1}}\right)^{N_A-1}\frac{\exp[-\sigma_A^{-1}\lambda]}{(\sigma_{Bj}^{-1}-\sigma_A^{-1})^{N_B-N+1}}\\
 &&+\frac{(N_A+N_B-N)!}{(N_B-N)!}\left(-\frac{\partial}{\partial \sigma_{Bj}^{-1}}\right)^{N_B-N}\frac{\exp[-\sigma_{Bj}^{-1}\lambda]}{(\sigma_A^{-1}-\sigma_{Bj}^{-1})^{N_A}}\nonumber\\ &=&\exp[-\sigma_A^{-1}\lambda]\sum_{k=0}^{N_A-1}(-1)^k\frac{(N_A+N_B-N)!(N_B-N+k)!}{k!(N_A-1-k)!(N_B-N)!}\frac{\lambda^{N_A-1-k}}{(\sigma_{Bj}^{-1}-\sigma_A^{-1})^{N_B-N+1+k}}\nonumber\\ &&+\exp[-\sigma_{Bj}^{-1}\lambda]\sum_{k=0}^{N_B-N}(-1)^k\frac{(N_A+N_B-N)!(N_A-1+k)!}{k!(N_B-N-k)!(N_A-1)!}\frac{\lambda^{N_B-N-k}}{(\sigma_A^{-1}-\sigma_{Bj}^{-1})^{N_A+k}}\,.\nonumber
\end{eqnarray}
Comparing this result with eq.~(\ref{jpdf}) derived in~\cite{Kumar:2014} we can read off the following identity for Kummer's confluent hypergeometric function
\begin{equation}\label{jpdf-funct} _1F_1(a;b;x)=\sum_{j=0}^{b-a-1}\frac{(-1)^a(b-1)!(a-1+j)!}{j!(b-a-1-j)!(a-1)!}\frac{1}{x^{a+j}}+\e^{x}\sum_{j=0}^{a-1}(-1)^j\frac{(b-1)!(b-a-1+j)!}{j!(a-1-j)!(b-a-1)!}\frac{1}{x^{b-a+j}}\,,
\end{equation}
for $b>a$ positive integers. This expression seems to be divergent at $x=0$. However one can check term by term that all negative powers in $x$ cancel in both sums. The existence of such expressions is well known for $a-b$ integer or $a$ a positive integer, cf.~\cite{NIST}, and can be derived alternatively using the recursion for $_1F_1(a;b;x)$ together with the known initial conditions for $_1F_1(a;a;x)=\exp[z]$ and for $_1F_1(a;a+1;x)$ in terms of the incomplete Gamma function.

\sect{Extension of Andr\'eief's Integration Formula}\label{Aint}

For completeness we quote here the generalisation of Andr\'eief's integration formula derived in \cite{Kieburg:2010}, as it is used several times in the main text. Let $R_j(x)$, $1\leq j\leq N+k$, and $S_j(x)$, $1\leq j\leq N+l$, be suitable integrable functions and $\{r_{ab}\}_{1\leq a\leq k}^{1\leq b\leq N+k}$ and $\{s_{ab}\}_{1\leq a\leq l}^{1\leq b\leq N+l}$ be two constant matrices. Then the following integral identity holds,
\bea
&&\prod_{j=1}^N\int dx_j
\det\left[
\begin{array}{r}
R_b(x_a)\big|_{1\leq a\leq N}^{1\leq b\leq N+k}\\
\\
r_{ab}\big|_{1\leq a\leq k}^{1\leq b\leq N+k}\\
\end{array}
\right]
\det\left[
\begin{array}{r}
S_b(x_a)\big|_{1\leq a\leq N}^{1\leq b\leq N+l}\\
\\
s_{ab}\big|_{1\leq a\leq l}^{1\leq b\leq N+l}\\
\end{array}
\right]
\label{genAnd}\\
&=& (-1)^{kl}N!\det\left[ 
\begin{array}{cc}
\int dx R_a(x) S_b(x)\big|_{1\leq a\leq N+k}^{1\leq b\leq N+l} 
& r_{ba}\big|^{1\leq b\leq k}_{1\leq a\leq N+k}
\\
&\\
s_{ab}\big|_{1\leq a\leq l}^{1\leq b\leq N+l}
& {\mathbf 0}_{l\times k}\\
\end{array}
\right].
\nn
\eea
In \cite{Kieburg:2010} complex integrals were considered while here we restricted the integration to real domains. Indeed one can replace the integrals by any linear functionals $R_j$ acting on the functions $S_j$ since eq.~\eqref{genAnd} is only an algebraic identity. It needs the linearity of the integral, the multi-linearity and the skew-symmetry of the determinant. Hence the identity~\eqref{genAnd} would also be true for sums or more complicated, in particular 
higher-dimensional integrals.

\sect{Expectation Value of Ratios of Characteristic Polynomials}\label{susyiso}

The approach to calculate $Z_{q|p}(X)$, see eq.~\eqref{Zpqdef}, works almost identically as the one in calculating the average of the single characteristic polynomial $P_N(x)$, see eq.~\eqref{OP}. The crucial modification enters due to the additional characteristic polynomial in the denominator. The source variables $y_l$ need an imaginary increment to regularise the integral. Thus, let us define $L={\rm diag}({\rm sign}\,{\rm Im}(y_1),\ldots,{\rm sign}\,{\rm Im}(y_q))$ that contains the signs of the imaginary parts of the source terms $y_j$. We assume that we have $q_+$ positive imaginary increments and $q_-$ negative ones, $q_++q_-=q$. Then the supersymmetric group involved in the partition function $Z_{q|p}(X)$ is ${\rm U}(q_+,q_-|p)$, see~\cite{Fyo,Zirn-noncom,KVZ}. Indeed we find this group again after mapping the average
\begin{equation}\label{Z-sdet}
Z_{q|p}(X)=\int [dA] \int [dB] \mathcal{P}_A(A)\mathcal{P}_B(B)\,{\rm Sdet}^{-1}\left(\eins_N\otimes X-(AA^\dagger+BB^\dagger)\otimes\eins_{q|p}\right)
\end{equation}
to superspace. We already rewrote the ratio of characteristic polynomials into the compact notation of a superdeterminant to be defined below, as a tensor of $N\times N$ ordinary matrices and of $(q|p)\times(q|p)$ supermatrices which we will introduce next.

\subsection{Brief introduction to the superalgebra of supermatrices}\label{sec:recall}

Let us briefly recall the crucial objects of superanalysis and superalgebra with supermatrices and introduce our notation. A more thorough introduction into this topic can be found in~\cite{Berezin}. Any complex rectangular supermatrix $\sigma$ of superdimensions $(q_1|p_1)\times (q_2|p_2)$ can be arranged into four matrix blocks,
\begin{equation}\label{susy-matrix-def}
\sigma=\left(\begin{array}{cc} \sigma_{\rm BB} & \sigma_{\rm BF} \\ \sigma_{\rm FB} & \sigma_{\rm FF}  \end{array}\right)\,.
\end{equation}
The $q_1\times q_2$ boson-boson block $\sigma_{\rm BB}$ and the $p_1\times p_2$ fermion-fermion block $\sigma_{\rm FF}$ comprise commuting variables, only. Note that this does not imply that they do not contain nilpotent terms, they indeed can. In contrast to the diagonal blocks the $q_1\times p_2$ boson-fermion block $\sigma_{\rm BF}$ and the $p_1\times q_2$ fermion-boson block $\sigma_{\rm FB}$ only consist of anti-commuting and, thus, nilpotent matrix entries. We denote by the set ${\rm Gl}(q_1|p_1;q_2|p_2)$ those supermatrices where the matrix entries of $\sigma_{\rm BB}$ and $\sigma_{\rm FF}$ are ordinary complex numbers and the matrix entries of $\sigma_{\rm BF}$ and $\sigma_{\rm FB}$ are independent complex Grassmann variables. In the case that $q=q_1=q_2$ and $p=p_1=p_2$ we abbreviate this set by ${\rm Gl}(q|p)$. Another important set we need is the coset ${\rm Herm}_+(q_+,q_-|p)={\rm Gl}(q_++q_-|p)/{\rm U}(q_+,q_-|p)$. A supermatrix $U\in{\rm Herm}_+(q_+,q_-|p)$ has a boson-boson block which is $L$-Hermitian, i.e. $U_{\rm BB}^\dagger=LU_{\rm BB}L$, and fulfils the positivity condition $LU_{\rm BB}>0$. We emphasise that these two properties imply that $U_{\rm BB}$ has $q_+$ positive real eigenvalues and $q_-$ negative real eigenvalues. The fermion-fermion block of $U$ is unitary, i.e. $U_{\rm FF}^\dagger=U_{\rm FF}^{-1}$ and the off-diagonal blocks consist of independent complex Grassmann variables with the condition $U_{\rm FB}^\dagger=U_{\rm BF}$. If $q_-=0$ we write ${\rm Herm}_+(q|p)={\rm Gl}(q|p)/{\rm U}(q|p)$. A matrix $V$ in the non-compact supergroup ${\rm U}(q_+,q_-|p)$ satisfies the identity
\begin{equation}\label{susy-unitary-def}
V^{-1}=\widehat{L}V^\dagger\widehat{L}\ ,
\end{equation}
with the diagonal supermatrix $\widehat{L}={\rm diag}(L,\eins_p)$.

Two important functions of a supermatrix $\sigma\in{\rm Gl}(q|p)$ are the supertrace,
\begin{equation}\label{susy-tr}
{\rm Str} \sigma=\Tr \sigma_{\rm BB}-\Tr \sigma_{\rm FF}\,,
\end{equation}
and the superdeterminant,
\begin{equation}\label{susy-det}
{\rm Sdet} (\sigma)=\frac{\det\left[\sigma_{\rm BB}-\sigma_{\rm BF}\sigma_{\rm FF}^{-1}\sigma_{\rm FB}\right]}{\det \left[\sigma_{\rm FF}\right]}=\frac{\det \left[\sigma_{\rm BB}\right]}{\det\left[\sigma_{\rm FF}-\sigma_{\rm FB}\sigma_{\rm BB}^{-1}\sigma_{\rm BF}\right]}\,,
\end{equation}
where here we need $\sigma_{\rm FF}$ and 
$\sigma_{\rm BB}$
to be invertible. The definitions are chosen in such a way that the properties of cyclic permutation invariance (${\rm Str}\, AB={\rm Str}\, BA$), or factorisation (${\rm Sdet}\, AB={\rm Sdet}\, A\,{\rm Sdet}\, B$), and the relation ${\rm ln}\,{\rm Sdet} (A)={\rm STr}\,{\rm ln}\,A$ are natural generalisations from the ordinary trace and determinant. Note that for the invariance under cyclic permutation the supermatrices $A$ and $B$ can also be rectangular while for the other properties we need square supermatrices.

\subsection{Mapping to superspace}\label{sec:calc}

To keep the calculation simple we omit the normalisation constant of $Z_{q|p}(X)$. In the end we fix it via the asymptotic
\begin{equation}\label{asymp}
\lim_{\epsilon\to\infty} \epsilon^{N(q-p)}Z_{q|p}(\epsilon X)=\Sdet^{-N}(X)\,.
\end{equation}

In a first step we rewrite the superdeterminant in eq.~\eqref{Z-sdet} as a Gaussian integral of a rectangular supermatrix $V\in{\rm Gl}(N|0;q|p)$, i.e.
\begin{equation}\label{sdet-Gauss}
{\rm Sdet}^{-1}(\eins_N\otimes X-(AA^\dagger+BB^\dagger)\otimes\eins_{q|p})=(-1)^{Nq_-}i^{N(q-p)}\frac{\int [dV] \e^{i\Tr  V\widehat{L}XV^\dagger-i\Tr(AA^\dagger+BB^\dagger)V\widehat{L}V^\dagger}}{\int [dV] \e^{-\Tr VV^\dagger}}\,.
\end{equation}
The prefactor comes from the factor $i\widehat{L}$ which has to be introduced into the superdeterminant to guarantee the positive definiteness of the Hermitian part of the matrix in the Gaussian integral. In this way the imaginary increment carries over to superspace. In particular, it allows us to interchange the integrals over $A$ and $B$ with those over $V$ since all integrals are absolutely integrable.

The integrals over $A$ and $B$ are purely Gaussian now, cf. eqs.~\eqref{PABdef} and~\eqref{sdet-Gauss}, such that we can perform them and find
\begin{equation}\label{Z-1}
Z_{q|p}(X)\propto\int [dV] \exp[i\Tr  V\widehat{L}XV^\dagger]\,{\det}^{-N_A}\left[\Sigma_A^{-1}+iV\widehat{L}V^\dagger\right]{\det}^{-N_B}\left[\Sigma_B^{-1}+iV\widehat{L}V^\dagger\right]\,.
\end{equation}
The two determinants are immediately regularised by the imaginary unit in front of $V\widehat{L}V^\dagger$.

In the next step we apply the duality relation
\begin{equation}\label{duality}
{\det}^{-N_A}\left[\Sigma_A^{-1}+iV\widehat{L}V^\dagger\right]={\det}^{N_A}\left[\Sigma_A\right] \Sdet^{-N_A}(\eins_{q|p}+i \widehat{L}V^\dagger\Sigma_AV)\,,
\end{equation}
and similarly for the determinant with $\Sigma_B$. This relation is true because one can identify $\det[\ldots]=\Sdet(\ldots)$ for boson-boson blocks and use the invariance under cyclic permutation, $\Sdet(\eins-YZ)=\Sdet(\eins-ZY)$, reminiscent of the cyclic permutation invariance of the supertrace. Then the partition function reads
\begin{equation}\label{Z-2}
Z_{q|p}(X)\propto\int [dV] \exp[i\Tr  V\widehat{L}XV^\dagger] \Sdet^{-N_A}(\eins_{q|p}+i \widehat{L}V^\dagger\Sigma_AV)\Sdet^{-N_B}(\eins_{q|p}+i \widehat{L}V^\dagger\Sigma_BV)\,.
\end{equation}
To rewrite the two superdeterminants again as Gaussian integrals, namely as
\begin{equation}\label{Gauss-back}
\Sdet^{-N_A}(\eins_{q|p}+i \widehat{L}V^\dagger\Sigma_AV)=\frac{\int [d\widehat{W}_A]\exp[-\Str \widehat{W}_A^\dagger(\eins_{q|p}+i \widehat{L}V^\dagger\Sigma_AV)\widehat{W}_A]}{\int [d\widehat{W}_A]\exp[-\Str \widehat{W}_A^\dagger\widehat{W}_A]}\,,
\end{equation}
with $\widehat{W}_A\in{\rm Gl}(q|p;N_A|0)$ and analogously for the one with $\Sigma_B$, we need to discuss if the Hermitian part of the boson-boson block of $\eins_{q|p}+i \widehat{L}V^\dagger\Sigma_AV$ is positive definite, in particular if the Hermitian part of $\eins_q+iLV_{\rm BB}^\dagger\Sigma_AV_{\rm BB}$ is positive definite. Since $V_{\rm BB}$ is an ordinary complex  rectangular $q\times N_A$ matrix we need to know if the matrix $K=LV_{\rm BB}^\dagger\Sigma_AV_{\rm BB}$ has complex eigenvalues. Thereby we note that $LK$ is Hermitian and positive definite. The $L$-Hermiticity of $K$ tells us that the eigenvalues of $K$ are either real or complex conjugate parts. Moreover, we can block-diagonalise $K=U^{-1}\Xi U$ by a non-compact unitary matrix $U\in{\rm U}(q_+,q_-)$, i.e. $U^{-1}=LU^\dagger L$. In the simplest case of $L=\diag(\eins_{q_+},-\eins_{q_-})$ the matrix $\Xi $ is of the form 
\begin{equation}\label{blockdiag}
\Xi =\left(\begin{array}{c|c|c} \diag(x_{1,1},\dots,x_{1,q_+-l}) & 0 & 0 \\ \hline 0 & \begin{array}{cc} \diag(x_{2,1},\dots,x_{2,l}) & \diag(y_{2,1},\dots,y_{2,l}) \\ -\diag(y_{2,1},\dots,y_{2,l}) & \diag(x_{2,1},\dots,x_{2,l}) \end{array} & 0 \\ \hline 0 & 0 & \diag(x_{3,1},\dots,x_{3,q_--l}) \end{array}\right),
\end{equation} 
with $x_{1,j},x_{2,j},x_{3,j},y_{2,j}\in\mathbb{R}$ and $l=1,\ldots,\lfloor\min(q_+,q_-)/2\rfloor$. The floor function $\lfloor\ldots\rfloor$ yields the largest integer smaller than or equal to its argument. The block diagonalisation with the structure~\eqref{blockdiag} was also discussed in \cite{KVZ} where the situation $L=\gamma_5=\diag(\eins_n,-\eins_{n+\nu})$ was considered. The situation of a general $L$ is related to the structure~\eqref{blockdiag} by a simple permutation of rows and columns such that the assumption $L=\diag(\eins_{q_+},-\eins_{q_-})$ is not a restriction at all. The positivity of $LK=LU^{-1}\Xi U=U^\dagger L\Xi  U$ carries over to a positivity condition of $L\Xi $. This implies two things. First, $\Xi $ and, thus $K$, has no complex conjugated pairs of eigenvalues, i.e. $l=0$. Second, the eigenvalues $x_{1,j}$ are positive and the eigenvalues $x_{3,j}$ are negative definite. Hence $K$ has a real spectrum and the real part of each eigenvalue of $\eins_{q}+i \widehat{L}V^\dagger\Sigma_AV=U^{-1}(\eins_q+i\Xi )U$ is equal to $1$. This discussion justifies the Gaussian integral~\eqref{Gauss-back} and renders the integral absolutely integrable.

Interchanging the integrals over $V$ with those over $\widehat{W}_A\in{\rm Gl}(q|p;N_A|0)$ and $\widehat{W}_B\in{\rm Gl}(q|p;N_B|0)$, we integrate over $V$ and find
\begin{equation}\label{Z-3}
Z_{q|p}(X)\propto\int [d\widehat{W}_A] \int [d\widehat{W}_B] \e^{-\Str \widehat{W}_A^\dagger\widehat{W}_A-\Str \widehat{W}_B^\dagger\widehat{W}_B} \Sdet^{-1}(\eins_N\otimes X-\Sigma_A\otimes \widehat{W}_A\widehat{W}_A^\dagger-\Sigma_B\otimes \widehat{W}_B\widehat{W}_B^\dagger)\,.
\end{equation}
In the final step we apply the superbosonisation formula~\cite{Efetov,Som07,BA07,LSZ08,KSG09} for $N_A,N_B\geq N\geq q$ to replace $\widehat{W}_A\widehat{W}_A^\dagger$ and $\widehat{W}_B\widehat{W}_B^\dagger$ by $U_A,U_B\in{\rm Herm}_+(q|p)$. This yields
\begin{eqnarray}\label{Z-4}
Z_{q|p}(X)&\propto&\int d\mu(U_A) \int d\mu(U_B) \e^{-\Str U_A-\Str U_B} \Sdet^{N_A}(U_A)\Sdet^{N_B}(U_B)\\
&&\times\Sdet^{-1}(\eins_N\otimes X-\Sigma_A\otimes U_A-\Sigma_B\otimes U_B)\,.\nonumber
\end{eqnarray}
The Haar measure $d\mu$ of the coset ${\rm Herm}_+(q|p)$ is explicitly given by~\cite{Som07,KSG09,Hua}
\begin{equation}\label{Haar-susy}
 d\mu(U)=(2\pi i)^{-p}\,\Sdet^{p-q} (U) [dU]\,,
\end{equation}
where $[dU]$ is again the flat measure, in particular the product of differentials of all independent matrix entries. We emphasise that there is no natural normalisation of the Haar measure on the coset ${\rm Herm}_+(q|p)$ when $pq>0$; namely the volume of the supergroup ${\rm U}(1|1)$ vanishes due to Cauchy-like integration theorems~\cite{Wegner,Efe,KKG09}. Hence we choose the normalisation by convenience since $p$ contour integrals are involved, see the next subsection.

\subsection{Calculation of the normalisation constant}\label{sec:const}

Due to the asymptotic~\eqref{asymp} the proportionality constant in eq.~\eqref{Z-4} is the inverse of the following integral
\begin{equation}\label{const-def}
C^{-1}=C_{N_A}^{-1}C_{N_B}^{-1}=\int d\mu(U_A)  \e^{-\Str U_A} \Sdet^{N_A}(U_A) \int d\mu(U_B)\e^{-\Str U_B}\Sdet^{N_B}(U_B)\,.
\end{equation}
Hence the two integrals in $U_A$ and $U_B$ factorise.

Let $n\in\mathbb{N}$. To calculate the integral
\begin{equation}
C_n^{-1}=\int d\mu(U)  \e^{-\Str U} \Sdet^{n}(U)=\int \frac{[dU]}{(2\pi i)^{p}} \e^{-\Str U} \Sdet^{n+p-q}(U)\,,
\end{equation}
with $U\in{\rm Herm}_+(q|p)$, we first split the supermatrix $U$ into four blocks as in eq.~\eqref{susy-matrix-def}, where $U_{\rm BB}\in{\rm Herm}_+(q)$ is Hermitian and positive definite and $U_{\rm FF}\in{\rm U}(p)$ is unitary.

Employing the second equality~\eqref{susy-det} for the superdeterminant we shift $U_{\rm FF}\to U_{\rm FF}+U_{\rm FB}U_{\rm BB}^{-1}U_{\rm FB}^\dagger$. The integrals over the complex Grassmann variables comprised in $U_{\rm FB}$ become Gaussian such that
\begin{equation}\label{const-1}
C_n^{-1}=\int_{{\rm Herm}_+(q)} [dU_{\rm BB}] \e^{-\Tr U_{\rm BB}} {\det}^{n-q}\left[U_{\rm BB}\right]\int_{{\rm U}(p)} \frac{[dU_{\rm FF}]}{(2\pi i)^{p}} \e^{\Tr U_{\rm FF}} {\det}^{q-n-p}\left[U_{\rm FF}\right]\,.
\end{equation}
The two remaining integrals are well-known from \cite{Mehta:2004,KSG09}. The non-compact integral is related to the Laguerre ensemble
\begin{eqnarray}\label{LUE}
 \int_{{\rm Herm}_+(q)} [dU_{\rm BB}] \e^{-\Tr U_{\rm BB}} {\det}^{n-q}[U_{\rm BB}]&=& \left(\prod_{l=1}^{q}\frac{\pi^{l-1}}{l!}\right) \prod_{j=1}^q\int_0^\infty d\lambda_j \lambda_j^{n-q}\e^{-\lambda_j} \Delta_q^2(\{\lambda_a\})\\
&=&\prod_{l=0}^{q-1}\pi^{l}(n-q+l)!\nonumber\,.
\end{eqnarray}
The prefactor in the first line is the volume of the coset ${\rm U}(q)/[{\rm U}^q(1)\times\mathbb{S}(q)]$ and in the second line we get an additional factor from the Selberg integral of Laguerre type~\cite{Mehta:2004}. The compact integral is an integral over the circular unitary ensemble
\begin{eqnarray}
 \hspace*{-0.5cm}\int_{{\rm U}(p)} \frac{[dU_{\rm FF}]}{(2\pi i)^{p}} \e^{\Tr U_{\rm FF}} {\det}^{q-n-p}[U_{\rm FF}]&=& \left(\prod_{l=1}^{p}\frac{\pi^{l-1}}{l!}\right) \prod_{j=1}^p\int_0^{2\pi} \frac{d\varphi_j}{2\pi} \e^{i(q-n)\varphi_j}\e^{\e^{i\varphi_j}} |\Delta_p(\{\e^{i\varphi_a}\})|^2\label{CUE}\\
&=&\prod_{l=0}^{p-1}\frac{\pi^{l}}{(n-q+l)!}\,.\nn
\end{eqnarray}
The contour integrals in the second step are of Selberg-type and were computed in~\cite{KSG09}. The combination of these results for the constants together with eq.~\eqref{Z-4} yield the result~\eqref{gen-res}.

\section{Computation of the Spectral Density in the General Setting}\label{sec:density-calc}

To compute the spectral density for general empirical matrices $\Sigma_A$ and $\Sigma_B$ we start from eq.~\eqref{gen-res} with $p=q=1$. We choose the following explicit coordinates for the supermatrices
\begin{equation}\label{susy-matrix-coord}
 U_{A/B}=\left(\begin{array}{cc} s_{A/B} & \eta_{A/B}^* \\ \eta_{A/B} & \e^{i\phi_{A/B}} \end{array}\right)\,,
\end{equation}
with $s_A,s_B\in\mathbb{R}_+$, $\phi_A,\phi_B\in[-\pi,\pi]$, and two complex Grassmann variables $\eta_A$ and $\eta_B$. The measure is given by
\begin{equation}\label{measure-expl}
d\mu(U_{A/B})=\frac{\e^{\imath\phi_{A/B}}
ds_{A/B}
d\phi_{A/B}d\eta_{A/B}d\eta_{A/B}^*}{2\pi}\,.
\end{equation}
To integrate over the Grassmann variables we have to expand the superdeterminants involved in eq.~\eqref{gen-res} which is
\begin{equation}\label{det-exp-1}
 \Sdet^l U_{A/B}=\left(\frac{s_{A/B}+\e^{-i\phi_{A/B}}\eta_{A/B}^*\eta_{A/B}}{\e^{i\phi_{A/B}}}\right)^l=s_{A/B}^l\e^{-i l\phi_{A/B}}\left(1+\frac{l}{s_{A/B}\e^{i\phi_{A/B}}}\eta_{A/B}^*\eta_{A/B}\right),
\end{equation}
and
\begin{eqnarray}\label{det-exp-2}
&&\Sdet^{-1}(\eins_N\otimes \diag(y,x)-\Sigma_A\otimes U_A-\Sigma_B\otimes U_B)\\
&=&\frac{\det(x\eins_N-\e^{i\phi_A}\Sigma_A-\e^{i\phi_B}\Sigma_B)}{\det(y\eins_N-s_A\Sigma_A-s_B\Sigma_B)}\biggl(1+\sum_{a,b\in\{A,B\}}\Tr(F_\phi\Sigma_aF_s\Sigma_b) \eta_{b}^*\eta_a\nonumber\\
&&+\Big(\Tr(F_\phi\Sigma_AF_s\Sigma_A)\Tr(F_\phi\Sigma_BF_s\Sigma_B)- \Tr(F_\phi\Sigma_AF_s\Sigma_B)\Tr(F_\phi\Sigma_BF_s\Sigma_A)\nonumber\\
&& -\Tr(F_\phi\Sigma_AF_s\Sigma_AF_\phi\Sigma_BF_s\Sigma_B) +\Tr(F_\phi\Sigma_AF_s\Sigma_BF_\phi\Sigma_BF_s\Sigma_A)\Big)\eta_A^*\eta_A\eta_B^*\eta_B\biggl).\nonumber
\end{eqnarray}
Here we have introduced the short-hand notation
\begin{equation}
F_\phi=(x\eins_N-\e^{i\phi_A}\Sigma_A-\e^{i\phi_B}\Sigma_B)^{-1}\ {\rm and}\ F_s=(y\eins_N-s_A\Sigma_A-s_B\Sigma_B)^{-1}.
\end{equation}
We insert these expressions into eq.~\eqref{gen-res} and integrate over the Grassmann variables,
\begin{eqnarray}
Z_{1|1}(y,x)&=&\int_{-\pi}^\pi\frac{d\phi_A}{2\pi}\int_{-\pi}^\pi\frac{d\phi_B}{2\pi}\int_0^\infty ds_A\int_0^\infty ds_B \e^{-s_A-s_B+\e^{i\phi_A}+\e^{i\phi_B}} s_A^{N_A}s_B^{N_B}\e^{i(1-N_A)\phi_A}\e^{i(1-N_B)\phi_B}\nonumber\\
&&\hspace*{-2.5cm}\times\frac{\det(x\eins_N-\e^{i\phi_A}\Sigma_A-\e^{i\phi_B}\Sigma_B)}{\det(y\eins_N-s_A\Sigma_A-s_B\Sigma_B)}\biggl(\frac{N_AN_B}{s_As_B\e^{i\phi_A}\e^{i\phi_B}}+\frac{N_B}{s_B\e^{i\phi_B}}\Tr(F_\phi\Sigma_AF_s\Sigma_A)+\frac{N_A}{s_A\e^{i\phi_A}}\Tr(F_\phi\Sigma_BF_s\Sigma_B)\nonumber\\
&&+\Big(\Tr(F_\phi\Sigma_AF_s\Sigma_A)\Tr(F_\phi\Sigma_BF_s\Sigma_B)- \Tr(F_\phi\Sigma_AF_s\Sigma_B)\Tr(F_\phi\Sigma_BF_s\Sigma_A)\nonumber\\
&& -\Tr(F_\phi\Sigma_AF_s\Sigma_AF_\phi\Sigma_BF_s\Sigma_B) +\Tr(F_\phi\Sigma_AF_s\Sigma_BF_\phi\Sigma_BF_s\Sigma_A)\Big)\biggl).\label{calc-app1}
\end{eqnarray}
The integration over the two angles can be summarised in terms of the 
expression
\begin{eqnarray}
&&\int_{-\pi}^\pi\frac{d\phi_A}{2\pi}\int_{-\pi}^\pi\frac{d\phi_B}{2\pi}\e^{-i\nu_A\phi_A}\e^{-i\nu_B\phi_B}\exp[\e^{i\phi_A}+\e^{i\phi_B}]\det(x\eins_N-\e^{i\phi_A}\Sigma_A-\e^{i\phi_B}\Sigma_B+S)\nonumber\\ &=&\frac{1}{\nu_A!\nu_B!}\left.\left(1+\frac{\partial}{\partial\mu_A}\right)^{\nu_A}\left(1+\frac{\partial}{\partial\mu_B}\right)^{\nu_B}\det(x\eins_N-\mu_A\Sigma_A-\mu_B\Sigma_B+S)\right|_{\lambda_{A/B}=0}\ ,\label{compact-int}
\end{eqnarray}
where $S$ is a fixed $N\times N$ matrix and $\nu_A,\nu_B\in\mathbb{N}_0$. Then the generating function~\eqref{calc-app1} can be written as
\begin{eqnarray}
&&Z_{1|1}(y,x)=\frac{1}{(N_A-1)!(N_B-1)!}\label{calc-app2}\\
&&\left(1+\frac{\partial}{\partial\mu_A}\right)^{N_A-1}\left(1+\frac{\partial}{\partial\mu_B}\right)^{N_B-1}\int_0^\infty ds_A\int_0^\infty ds_B \frac{\e^{-s_A-s_B} s_A^{N_A}s_B^{N_B}}{\det^3(y\eins_N-s_A\Sigma_A-s_B\Sigma_B)}\nonumber\\
&&\times\biggl(\frac{1}{s_As_B}\left(1+\frac{\partial}{\partial\mu_A}\right)\left(1+\frac{\partial}{\partial\mu_B}\right)+\frac{1}{s_B}\left(1+\frac{\partial}{\partial\mu_B}\right)\frac{\partial^2}{\partial z_A\partial z_A^*}+\frac{1}{s_A}\left(1+\frac{\partial}{\partial\mu_A}\right)\frac{\partial^2}{\partial z_B\partial z_B^*}\nonumber\\
&&+\left.\frac{\partial^4}{\partial z_A\partial z_A^*\partial z_B\partial z_B^*}\biggl)\det\left[\begin{array}{c|c} x\eins_N-\mu_A\Sigma_A-\mu_B\Sigma_B & \begin{array}{cc} \sqrt{2}z_A\Sigma_A\qquad & \sqrt{2}z_B\Sigma_B \end{array} \\ \hline \begin{array}{c} \sqrt{2}z_A^*\Sigma_A \\ \sqrt{2}z_B^*\Sigma_B \end{array} & \displaystyle\frac{y\eins_N-s_A\Sigma_A-s_B\Sigma_B}{2} \otimes\left(\begin{array}{cc} 1 & i \\ i & 1 \end{array}\right) \end{array}\right]\right|_{\mu_{A/B}=z_{A/B}=0}
\hspace*{-0.75cm}.
\nonumber
\end{eqnarray}
Here
we rewrote the integration over the two angles and the Grassmann variables as derivatives with respect to the source variables $\mu_{A/B}$ and $z_{A/B}$. Taking the derivative with respect to $x$, setting $x=y$ and taking the imaginary part in the limit ${\rm Im}(y)\to0^+$ yields the result~\eqref{level-result}.

\end{appendix}


\end{document}